\documentclass[aps,prx,twocolumn,superscriptaddress]{revtex4-2}
\usepackage[a4paper,margin=1.25cm]{geometry}
\usepackage{times}
\usepackage{graphicx}
\usepackage[font={small}]{caption}
\usepackage{subcaption}
\usepackage{amsmath}
\usepackage{amssymb}
\usepackage{tabularx}
\usepackage{placeins}
\usepackage{units}
\usepackage{url}
\usepackage{multirow}
\usepackage{braket}
\usepackage{color}
\usepackage{hyperref}
\usepackage[normalem]{ulem}
\usepackage{breakcites}
\usepackage{adjustbox}
\usepackage{mathtools}
\usepackage{float}
\usepackage{adjustbox}
\usepackage{soul}
\usepackage{dcolumn}
\usepackage{bm}
\usepackage{comment}
\usepackage{amssymb}
\usepackage{nameref}
\usepackage{hyperref}
\usepackage[english]{babel}

\usepackage[dvipsnames]{xcolor}

\begin{document}

\title{A new generation of effective core potentials: Selected heavy $5d$ and $6p$ elements}

\author{Omar Madany}
\email{osmadany@ncsu.edu}
\author{Lubos Mitas}
\affiliation{Department of Physics, North Carolina State University, Raleigh, North Carolina 27695-8202, USA}

\begin{abstract}
We expand the correlation-consistent effective core potentials (ccECPs) library by developing semi-local pseudopotentials and matching basis sets by heavy-elements from $5d$ (Hf, Os, Hg) and $6p$ (Tl, Po, At, Rn) blocks. In order
to accurately capture scalar relativistic effects, spin-orbit coupling, and electron-electron correlation, we implement a tiered core-valence partitioning strategy across three distinct resolutions. 
This includes a small 60-core (Hf, Os, Hg) that explicitly correlates subvalence shells, a large 78-core definition for the main-group elements that rigorously accounts for core polarization and relaxation effects in sparse valence environments, and an intermediate 68-core partition for Hg and Tl.
This 68-core architecture represents a unique development in the ccECP library, optimizing the balance between accuracy and computational efficiency in a manner unexplored for ligther elements. 
Optimized against relativistic all-electron CCSD(T) references, the ccECPs deliver outstanding atomic precision, achieving a global average atomic low-lying states deviation of just 0.045 eV. 
This accuracy translates directly to robust molecular transferability, systematically restricting dissociation energy discrepancies to under 0.03 eV, equilibrium bond lengths to within 0.005 \AA. 
By enforcing a regularized, finite potential at the origin for enhanced numerical stability in stochastic quantum Monte Carlo methods, this library removes a critical methodological bottleneck for predictive many-body simulations of heavy-element systems and materials.

\end{abstract}

\maketitle

\section{Introduction}
Within the last decade, the development of correlation-consistent effective core potentials (ccECPs) has sought to modernize the landscape of pseudopotentials\cite{Bennett2017,Bennett2018,Annaberdiyev2018,Wang2019,Wang2022,Kincaid2022,annaberdiyev2023,Haihan2024,Omar2025,pseudopotentiallibrary}. One central goal in the ccECP development is substituting chemically inert core electrons and concentrate computational efforts on the valence subspace without introducing accuracy bottlenecks and contain discrepancies to 1 kcal/mol, meeting the threshold of chemical accuracy required for predictive many-body simulations. By shifting the construction paradigm from one-particle mean-field frameworks such as Dirac-Hartree-Fock (DHF)\cite{hartree1928,fock1930,bertha1935} or density functional theory\cite{kohnsham1965} methods to explicit many-body, leveraging coupled-cluster (CC)\cite{coester1958,cizek1966} and configuration interaction (CI)\cite{boys1950,shavitt1977} references while embedding Coulomb cusp corrections advantageous for quantum Monte Carlo (QMC) methods\cite{foulkes2001,acc_engI,acc_engII}, lead the ccECP library to emerge as a tested choice for ECPs in the community\cite{ph1,ph2}.

The efficacy of ccECPs in providing a systematically improvable and relativistic description of the valence manifold has led to their broad adoption across diverse landscapes of electronic structure theory. In the active area of neural network wavefunctions and machine learning, ccECPs have become the standard for incorporating heavy elements into periodic architectures\cite{xiang2022} and molecular suites\cite{zeno2023}, facilitating high-accuracy assessments of pseudopotential performance \cite{mengsa2024}, excited-state vertical energies\cite{bernat2024}, spin-symmetry enforcement in biradicals \cite{Zhe2024}, and the acceleration of many-body force estimation for large-scale systems \cite{cancan2023, Michael2025}.
For bulk solids and periodic materials, ccECPs have been instrumental in resolving the subtle energetics in silicon\cite{gani2021}, magnetic delafossite oxides\cite{hyeondeok2024}, boron nitride polymorphs\cite{yutaka2022}, and NiTi shape-memory alloys\cite{kevin2026}. This accuracy extends to characterizing defect resistance and magnetic ordering in complex chalcogenides\cite{kayahan2024, kayahan2025, jeonghwan2023} and layered magnets such as $\alpha$-RuCl$_3$\cite{gani2022} and $\alpha$-Fe \cite{gani2024}, and characterizing complex perovskites\cite{cody2020}. 
In two-dimensional materials and surface science, these potentials have unraveled sensitive electronic gaps in fluorographene\cite{matus2020}, interlayer coupling in twisted bilayers\cite{jeonghwan2024}, and the topological properties of heterostructures\cite{daniel2023, daniel2024}, alongside physisorption and chemisorption studies on metal-decorated graphene\cite{marketa2023, yasmine2023}. 
The predictive power of this framework extends from periodic surfaces to discrete molecular complexes and non-covalent regimes. Applications to molecular systems and non-covalent interactions have achieved sub-chemical accuracy in biological ligand-pocket binding\cite{mirela2025}, large complexes\cite{kousuke2026, anouar2020} and, halogenation effects\cite{ritaj2023}. This precision is further demonstrated in the rigorous characterization of radical thermochemistry\cite{timothy2024, gopal2024}, nitrogenase model complexes\cite{victor2024}, thiolated gold nanoclusters\cite{juha2023}, and metallophthalocyanine spectra\cite{qunfei2021}. 
Finally, the ccECP library provides the high-fidelity reference data required for large-scale synthetic dataset construction \cite{anouar2025, danish2025, maxwell2026} and serves as a foundational component in advanced software packages \cite{paul2020, william2023, kousuke2023, kousuke2020}. It has been applied in community-wide reproducibility benchmark studies across independent codes\cite{flaviano2025}, cross-validations of DMC and auxiliary-field QMC\cite{fionn2020} and the scaling of many-body ionic force estimators for heavier elements\cite{juha2021}.
By completing the systematic modernization of the 5d and 6p elements, this work extends the predictive reach of this framework to the heaviest regions of the periodic table.

The electronic structure of the $5d$ and $6p$ elements is governed by the lanthanide contraction and pronounced relativistic effects, where the magnitude of scalar contraction and SO coupling is comparable to the valence correlation energy. In this high nuclear charge regime, these effects are not merely perturbative; they fundamentally redefine the spatial distribution and energetic ordering of the valence orbitals. This redistribution of electron density blurs the boundary between core and valence, rendering the frozen-core approximation highly susceptible to deviations arising from the high polarizability of the outer semi-core states in diverse coordination geometries.

We address the distinct electronic and relativistic of these selected elements through a systematic core-valence partitioning strategy operating across three distinct resolutions. For the heavy $5d$ transition metals, we implement a definition that explicitly retains the $5s$ and $5p$ subvalence shells to capture profound core-valence correlation effects. To optimally balance chemical fidelity with computational efficiency, we introduce a partition for mercury and thallium that absorbs the inner subvalence space while continuing to explicitly correlate the $5d$ manifold. This intermediate resolution represents a unique architectural development in the ccECP library; lighter counterparts in the $3d/4p$ and $4d/5p$ rows bypassed this tier entirely, relying instead on either expansive small-core formulations (such as 20-electron Zn/Cd and 21-electron In) or minimal large-core treatments (such as 3-electron Ga/In). Finally, a large-core definition is applied to the $6p$ main-group elements, capitalizing on the increasingly localized nature of their filled $5d$ shells. Because this large-core definition restricts the active space to a small number of valence electrons (down to 3 in Tl), capturing the residual core physics becomes exceptionally critical to avoid introducing significant deviations as the core undergoes considerable polarization and relaxation in response to changes in the coordination environment. This tiered framework consistently recovers critical physical observables, accurately reproducing the structural and energetic trends driven by scalar relativistic contraction and strong spin-orbit coupling across the entire row.

Beyond energetic accuracy, the ccECP formalism relies on specific mathematical regularizations. A key feature of our construction is the bounded character of corresponding potentials in all channels, including the local channel. By enforcing a finite value and vanishing first derivative at the origin, the potentials simplify the description of orbitals for various methodologies while significantly suppressing energy fluctuations in high-precision stochastic many-body methods. Also, we carry out conversion from semi-local to non-local separable Kleinman-Bylander forms, enabling straightforward application within plane-wave computational frameworks. Validation efforts extend across diverse chemical environments, utilizing hydride and oxide binding curves to evaluate the balance between electronic polarization and Pauli repulsion across a wide range of interatomic distances.

This completion of the ccECP library for the Cs--Rn row provides the community with high-accuracy valence Hamiltonians for the heaviest elements of the periodic table. By integrating core–core and core–valence correlations with relativistic effects into a systematically derived ECPs, we provide a reliable foundation for predictive many-body modeling of complex heavy-atom molecular systems and materials.

\section{Methods}
\label{Methods}
\subsection{Effective Core Potential Formalism}
\label{ECP form}
The relativistic valence electronic structure is treated within a two-component spinor framework\cite{Lee-1977,Ermler1981}. Following the successful form implemented in previous ccECPs\cite{Wang2022,Kincaid2022,annaberdiyev2023,Haihan2024,Omar2025}, the total semi-local SO relativistic ECP (SOREP) is partitioned into a scalar, spin-averaged component and a vector SO coupling term:
\begin{equation}\label{eq:sorep_arep_so}
    V^{\text{SOREP}}_i = V^{\text{AREP}}_i + V^{\text{SO}}_i,
\end{equation}

The scalar average relativistic effective core potential (AREP) captures the effective nuclear field experienced by valence electrons with scalar relativistic effects included. It is constructed from a local potential ($V_{loc}$) and non-local projectors ($\Delta V_\ell$) acting on specific angular momentum subspaces:
\begin{equation}\begin{aligned}\label{Varepform}
    V^{\text{AREP}}_i &= V_{loc}(r) 
    + \sum_{\ell=0}^{\ell_{\text{max}}=L-1} \Delta V_\ell^{\text{AREP}}(r) \\
    &\quad \times \sum_{m=-\ell}^{\ell} | \ell m \rangle \langle \ell m |,
\end{aligned}\end{equation}

where $\Delta V_\ell^{\text{AREP}}(r) = V_\ell - V_{loc}$ represents the non-local radial correction for the $\ell$-th channel, and $L$ is the angular momentum of the local channel, chosen to be $\ell_{\text{max}}+1$.

The full relativistic potential is reconstructed by adding to the scalar AREP a net SO potential ($\Delta V_{\ell}^{\text{SO}}$) and the corresponding projection of the spin angular momentum onto the orbital momentum:
\begin{equation}
    V^{\text{SO}}_i = \sum_{\ell=1}^{\ell_{\text{max}}} \Delta V_{\ell}^{\text{SO}} \, P_{\ell} \, \vec{\ell} \cdot \vec{s} \, P_{\ell}
\end{equation}

To strictly enforce a finite potential at the origin and its zero derivative  the local potential $V_{loc}(r)$  is parameterized as:
\begin{equation}
    \label{eq:AREP_local}
    \begin{aligned}
        V_{loc} (r) =& - \frac{Z_{\text{eff}}}{r} \left( 1 - e^{-\alpha_{L 1} r^2} \right) + \alpha_{L 1} Z_{\text{eff}} r e^{-\alpha_{L 2} r^2} \\
        &+ \sum_{k}^{} \beta_{L k} e^{-\alpha_{L k} r^2}
    \end{aligned}    
\end{equation}

Here, $Z_{\text{eff}}$ denotes the effective core charge. The analytic structure of the leading terms enforces a finite, smooth behavior at the origin (where $\nabla V_{L}|_{0}=0$), thereby regularizing the Coulomb singularity\cite{Ivan2001,BFD-2007,BFD-2008}.

Finally, the scalar $\Delta V_\ell^{\text{AREP}}$ and spin-orbit $\Delta V_\ell^{\text{SO}}$ non-local radial potentials are expanded using a standard Gaussian functions:
\begin{equation}\label{eq:AREP_non-local}
    \Delta V_\ell(r) = \sum_{k=1} \beta_{\ell k} r^{n_{\ell k} - 2} e^{-\alpha_{\ell k} r^2}
\end{equation}
where the set {$\alpha, \beta$} constitute the optimized variational space, spanning both Gaussian exponents and amplitudes while $n_{\ell k}$ are fixed integer powers. 

\subsection{Parameterization and Objective Function}
\label{sec:optpro}
The construction of the effective Hamiltonian follows a hierarchical optimization workflow designed to maximize both spectral accuracy and transferability. The process is anchored by high-fidelity all-electron (AE) benchmarks generated using the tenth-order Douglas-Kroll-Hess Hamiltonian within CC theory with singles, doubles, and perturbative triples (CCSD(T)) implemented in \textsc{MOLPRO} package\cite{DKH-2004,molpro-2012,molpro-2020,molpro-3,molpro-cc,molpro-seward,molpro-mcscf1,molpro-mcscf2}. To ensure that the reference data are unbiased by basis set artifacts, uncontracted AE Gaussian basis sets\cite{basis-5d-2005,basis-mdfstu-5d-1-2009,basis-6p-2015} are validated against numerical DHF\cite{yoon-DHF,Kotochigova1997} to minimize linear dependencies and provide the necessary augmentation terms.

The optimization is partitioned into two stages: first, establishing the scalar-relativistic AREP framework, followed by the SO contributions. Initial parameters are seeded from legacy potentials or prior ccECPs, explicitly modified to enforce the finite potential and zero-slope conditions at the nucleus required by the ccECP formalism.

The ccECP parameters were refined using sequential quadratic programming \cite{Spellucci-1998, Spellucci-1998-sqp}, often times to navigate the parameter space and mitigate local minima issues, we explored machine learning-based Bayesian optimization\cite{optuna2019} using the Tree-structured Parzen Estimator algorithm \cite{tpe2011}. The scalar parameters are determined by minimizing a hybrid objective function $\mathcal{O}^{2}(\xi)$ which comprises atomic spectral gaps, single-particle eigenvalues, and molecular binding energies. The function is defined as a thresholded weighted least-squares functional:
\begin{equation}\label{eq:obj_fxn}
    \mathcal{O}^{2}(\xi) = \sum_{k \in \Omega} \omega_k \left [ \delta_{k}(\xi) \cdot \mathbb{I}\left(|\delta_{k}(\xi)|>\tau \right) \right]^{2}
\end{equation}
where $\xi$ represents the set of Gaussian parameters, and $\Omega$ denotes the composite set of benchmarks. The discrepancy $\delta_{k}$ between the ECP and AE reference for observable $k$ is given by:
\begin{equation}
    \delta_{k}(\xi) = \Delta E_k^{\text{ECP}}(\xi) - \Delta E_k^{\text{AE}}(\xi)
\end{equation}
Here, $\omega_{k}$, is a scalar weight prioritizing equilibrium properties and low-lying transitions. The hard-threshold filter ensures that any observable converged within the chemical accuracy tolerance contributes zero penalty, preventing overfitting to numerical noise.

Upon convergence of the scalar potential, the SO terms are optimized using a similar objective function, targeting multiplet splitting energies obtained at the complete open-shell configuration interaction theory level using exact two-component Hamiltonian implemented in the \textsc{DIRAC} package\cite{DIRAC22,DIRAC22-2020,DIRAC-X2C, DIRAC-X2C-2}. Finally, consistent basis sets are generated and ccECPs are converted to grid-based radial XML formats and UPF Kleinman-Bylander projected versions using a modified \textsc{OPIUM} package\cite{Wang2022,opium}. The energy cutoffs required for plane-wave applications were assessed using \textsc{QUANTUM ESPRESSO}\cite{Giannozzi_2009, Giannozzi_2017, Giannozzi_2020} automated through \textsc{NEXUS} management system program\cite{nexus}. Detailed procedures regarding basis set construction and plane-wave cutoff testing are provided in the Supplementary Material.

\section{Results}
\label{Results}
\label{sec:results}
\begin{figure}[!htbp]
\centering
\includegraphics[width=1.00\columnwidth]{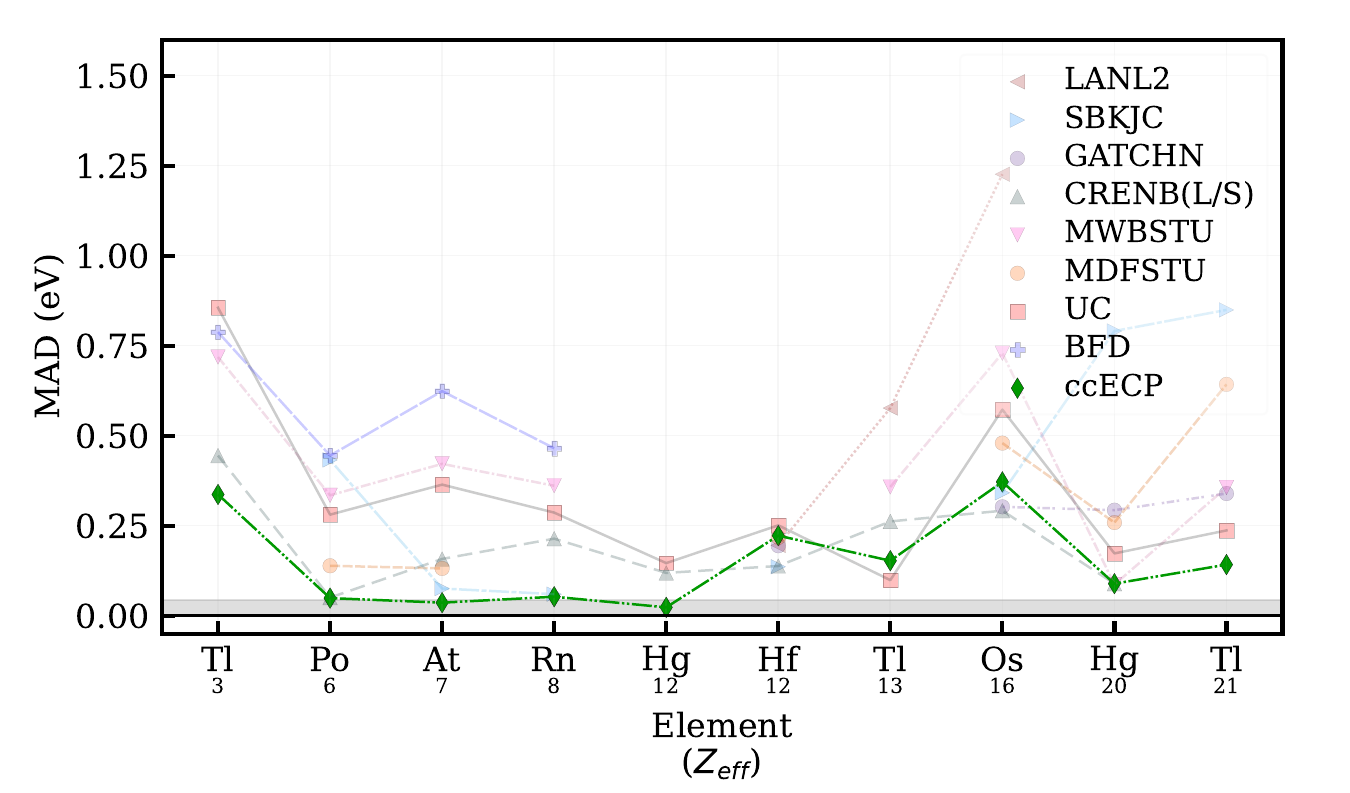}
\caption{Mean absolute deviation of the atomic excitation energies at the CCSD(T) level for various ECPs relative to the scalar relativistic all-electron reference. The performance of the ccECPs is compared against available legacy ECPs and frozen-core UC methods. The x-axis labels denote the element, and the number of valence electrons explicitly treated in the calculation.}
\label{fig:MAD_in_elements}
\end{figure}

In order to rigorously quantify the spectral properties of the constructed ccECPs, we use three primary statistical metrics evaluated against relativistic AE CCSD(T) reference data. The first, mean absolute deviation (MAD, Fig. \ref{fig:MAD_in_elements}), provides a global assessment of the energetic consistency across the entire sampled set of $N$ atomic states, defined as:
\begin{equation}
    \text{MAD} = \frac{1}{N} \sum_{s=1}^{N} \left| \Delta E^{\text{ECP}}_s - \Delta E^{\text{AE}}_s \right|
\end{equation}

\begin{figure}[!htbp]
\centering
\includegraphics[width=1.00\columnwidth]{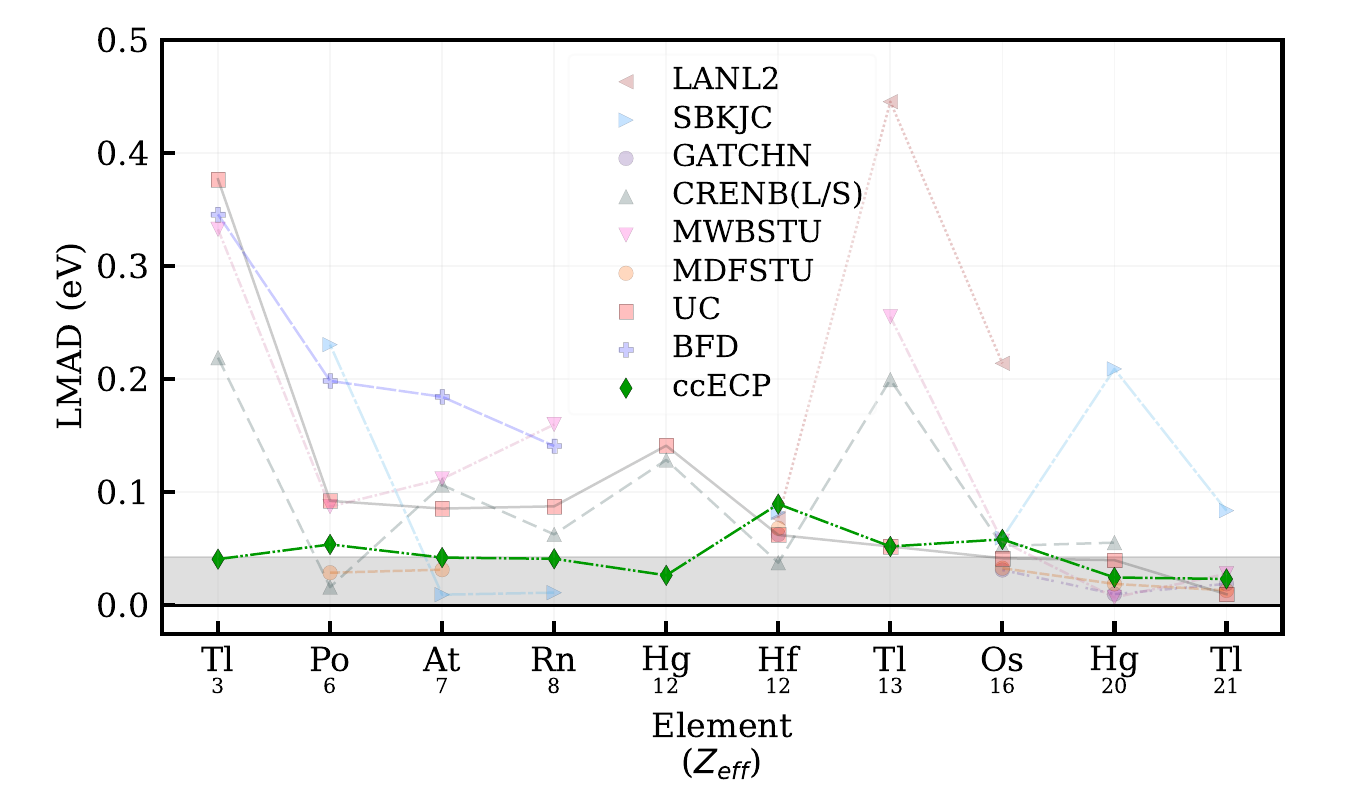}
\caption{
Low-lying mean absolute deviation of the atomic excitation energies at the CCSD(T) level. The ECP comparison scheme follow the same conventions detailed in Fig. \ref{fig:MAD_in_elements}). 
}
\label{fig:LMAD_in_elements}
\end{figure}

For the valence excitations most critical to properly capturing the electronic properties of valence, we utilize the low-lying mean absolute deviation (LMAD, Fig. \ref{fig:LMAD_in_elements}). This metric focuses on a specific subset $n$ comprising the electron affinity (where stable anions exist), the first and second ionization potentials, and the lowest-lying neutral excitations:
\begin{equation}
    \text{LMAD} = \frac{1}{n} \sum_{s=1}^{n} \left| \Delta E^{\text{ECP}}_s - \Delta E^{\text{AE}}_s \right|
\end{equation}

\begin{figure}[!htbp]
\centering
\includegraphics[width=1.00\columnwidth]{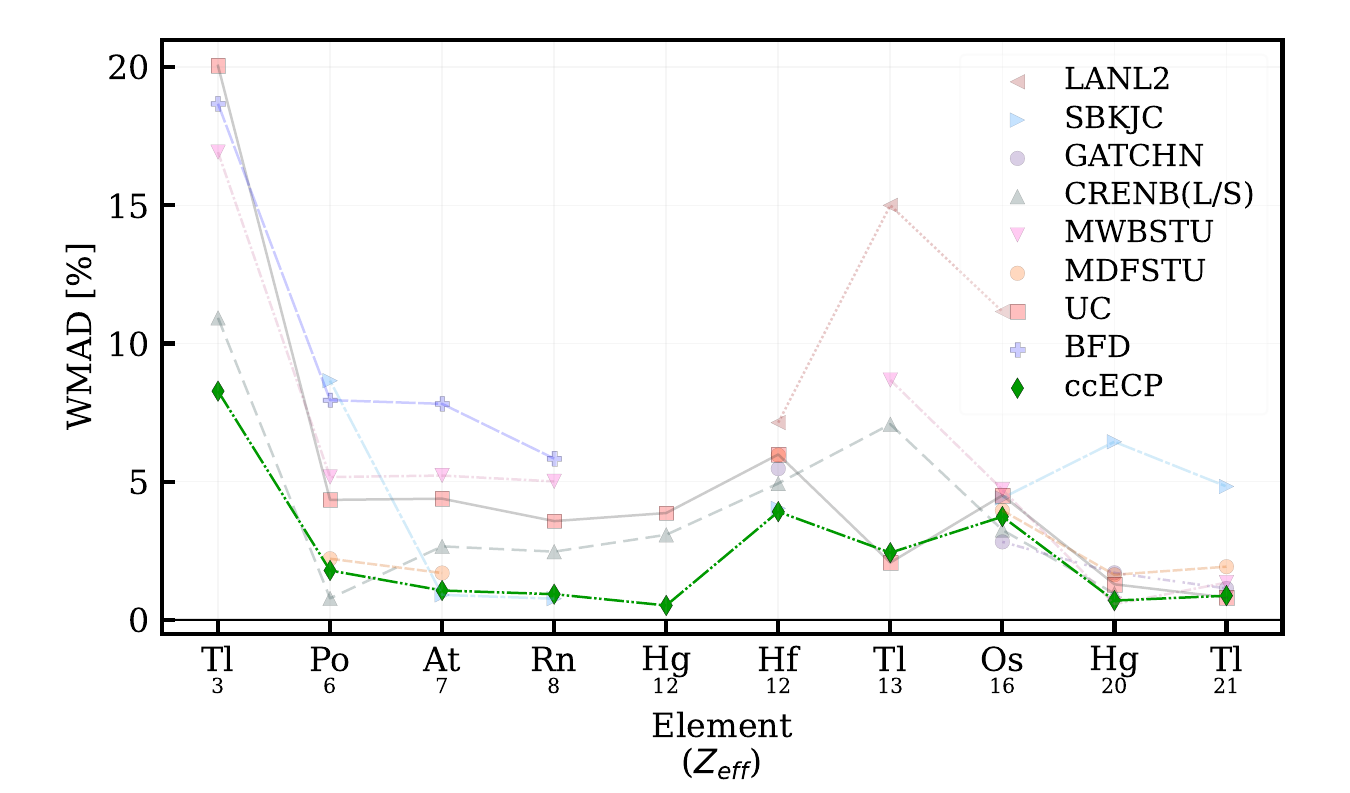}
\caption{
Weighted mean absolute deviation of the atomic excitation energies at the CCSD(T) level. The ECP comparison scheme follow the same conventions detailed in Fig. \ref{fig:MAD_in_elements}).
}
\label{fig:WMAD_in_elements}
\end{figure}

Finally, recognizing that absolute discrepancy magnitudes naturally scale with total transition energy, we introduce the weighted mean absolute deviation (WMAD, Fig. \ref{fig:WMAD_in_elements}). This metric normalizes the discrepancy by the inverse square root of the reference energy gap, providing a relative percentage discrepancy that balances contributions from high-energy core excitations and low-energy valence states:
\begin{equation}
    \text{WMAD} = \frac{1}{N} \sum_{s=1}^{N} \frac{100}{\sqrt{\left| \Delta E^{\text{AE}}_s \right|}}  \left| \Delta E^{\text{ECP}}_s - \Delta E^{\text{AE}}_s \right|
\end{equation}

\begin{figure}[!htbp]
\centering
\includegraphics[width=1.00\columnwidth]{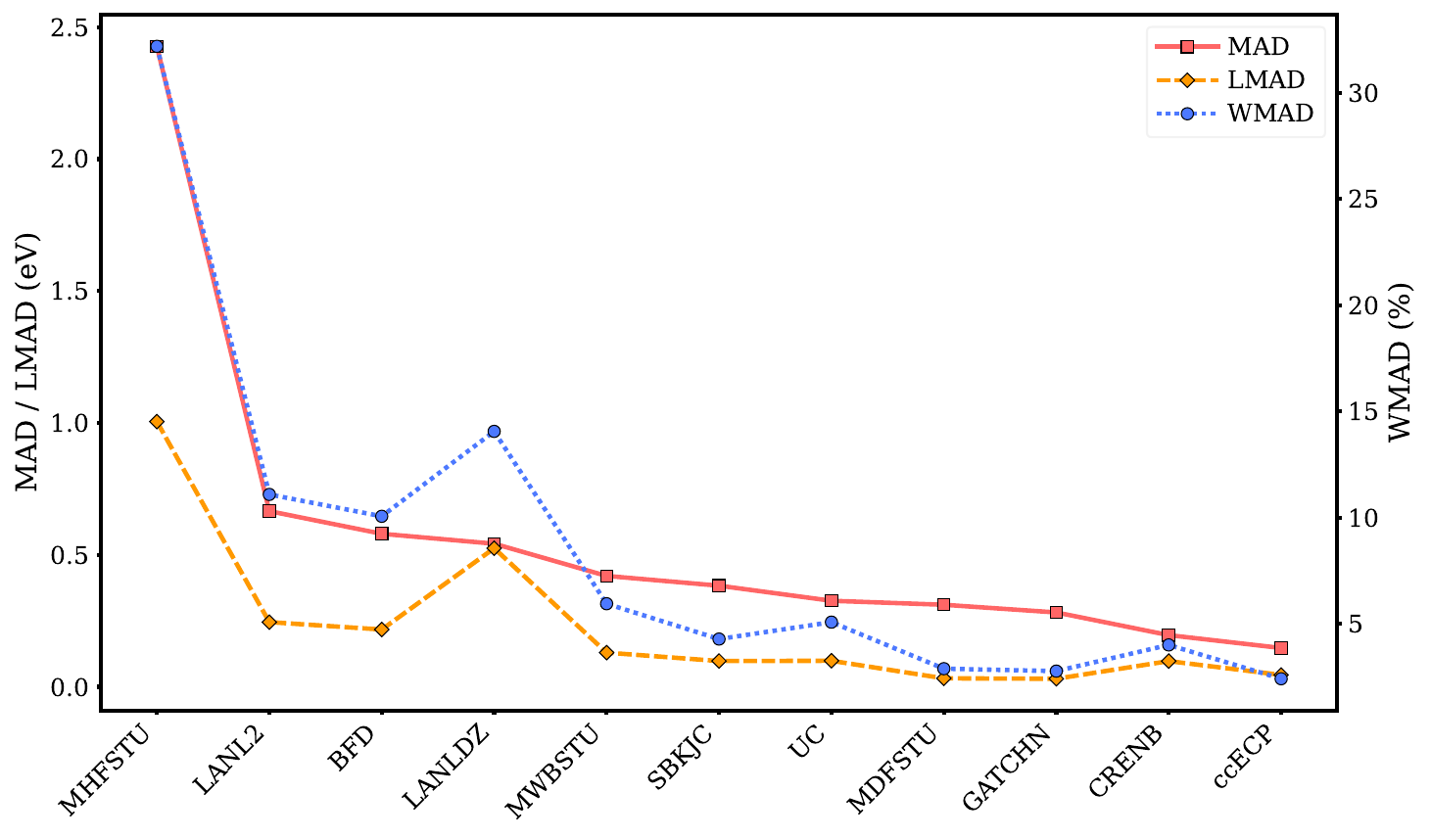}
\caption{Overall average atomic discrepancies (MAD, LMAD, and WMAD) for all evaluated $5d$ and $6p$ element pseudopotentials relative to relativistic AE CCSD(T) reference data. The primary (left) y-axis tracks absolute deviations (MAD/LMAD), while the secondary (right) y-axis charts the energy-normalized weighted deviation (WMAD).
}
\label{fig:summary_w_l_mad}
\end{figure}

To synthesize these atomic benchmarks across all evaluated elements and core partitions, the overall average discrepancies are compiled in Fig. \ref{fig:summary_w_l_mad}. The ccECP formalism demonstrates exceptional systematic accuracy, consistently outperforming available legacy pseudopotentials, and even the AE frozen-core/uncorrelated-core (UC)  reference across all metrics. In particular, ccECPs strictly bound the chemically critical low-lying excitations to an average LMAD of just $\sim$0.043~eV, effectively eliminating the severe valence discrepancies present in standard legacy potentials. Coupled with a highly consistent average WMAD of 2.44~\%, these metrics confirm the robustness and reliable transferability of the rigorously correlated ccECP framework.

Beyond scalar relativistic effects, we validate the SO components of the ccECPs by evaluating J-splitting patterns across a manifold of low-lying atomic states. The accuracy of the SO parameterization is quantified by the J-splitting mean absolute deviation (JMAD) relative to the relativistic AE benchmarks. This metric ensures that the components of the SO potential correctly capture the splitting of degenerate levels into their respective $J$ total angular momentum states. Comprehensive details are provided in the Supplementary Material.

We probed the robustness of the derived ccECPs beyond the atomic optimization limit by examination of  their transferability across the potential energy surfaces for representative hydride and oxide molecules. These systems function as an excellent testing set to investigate distinct interaction regimes, from the strong repulsion at very short bond distances (simulating thus bond lengths corresponding to high-pressure condensed phases) to the asymptotic dissociation limit. The fidelity of the ccECP description is quantified by the deviation of the binding energy ($\Delta(r)$) relative to the relativistic AE CCSD(T):
\begin{equation}
    \Delta(r) = D^{\text{ECP}}(r) - D^{\text{AE}}(r)
\end{equation}

In addition to these energy profiles, we also calculate a set of discrete spectroscopic observables, in particular, the dissociation energy ($D_e$), equilibrium bond length ($r_e$), the curvature parameter ($a$), and the vibrational frequency ($\omega_e$) via Morse potential fitting (see Supplementary material for details).

These molecular metrics form the basis of our comparative analysis against a comprehensive suite of previous ECPs--including BFD\cite{BFD-2007,BFD-2008}, MDFSTU\cite{basis-mdfstu-5d-1-2009,mdfstu-5d-2,mdfstu-6p-1,mdfstu-6p-2}, MWBSTU\cite{mwbstu-5d-1,mwbstu-6p-1,mwbstu-6p-2}, CRENB(S/L)\cite{CRENB-5d}, SBKJC\cite{sbkjc-5d}, GATCHN\cite{gatchn1997,gatchn2005,gatchn2007} and LANL2\cite{lanl2-5d} obtained through Basis Set Exchange\cite{pritchard2019a,feller1996a,schuchardt2007a}--alongside the UC method, namely only correlating the valence space in the AE case.

It is instructive to distinguish the theoretical footing of the ccECPs within this set. Unlike most previous parameterizations (with the exception of BFD), the ccECPs formalism explicitly regularizes the potentials, enforcing a finite potential at the nucleus, yielding finite, smooth potential functions that require no ad hoc cusp corrections. Crucially, this rigorous boundary condition is satisfied while simultaneously constraining Gaussian exponents to mitigate potential hardness and ensure suitability for plane-wave calculations. 
Note that singularities (effective charge Coulomb or $1/r^2$) in many previously constructed ECPs do not provide anything important for the valence properties to be reproduced correctly and accurately. Consequently, the results presented herein represent the optimal balance of accuracy achievable when balancing numerical convenience with high-precision many-body physics.

Furthermore, to ensure broad utility, we evaluated the computational efficiency of these potentials in plane-wave applications using their Kleinman-Bylander transformed variants. Across the library, we find that kinetic energy cutoffs of 40--60 Ry for the large-core [[Xe]$4f^{14}5d^{10}$] ccECPs, 200 Ry for the medium-core [[Xe]$4f^{14}$] ccECPs and 320--400 Ry for the small-core [[Kr]$4d^{10}4f^{14}$] ccECPs are sufficient to converge total energies to within 1 meV/electron relative to a high-cutoff reference (see Supplemental Material).

\subsection{Selected $5d$ elements}
For the selected $5d$ transition metals, we employ a consistent small-core partitioning defined by a [[Kr]$4d^{10}4f^{14}$] core. This approach treats the $5s$, $5p$, $5d$, and $6s$ shells explicitly, ensuring that semi-core polarization and core-valence correlation effects critical for this row are captured. Overall, the ccECPs for this series demonstrate superior accuracy in atomic spectral benchmarks compared to legacy ECPs and frozen-core approximations, yielding significantly lower discrepancies in high-energy excitations and fine-structure splittings. This spectral precision translates to robust molecular transferability, where the ccECPs accurately reproduce equilibrium properties across covalent and ionic systems.
\subsubsection{Hf}
The efficacy of the hafnium (Hf) ccECP ($Z_{\text{eff}}=12$) is demonstrated in the atomic spectrum benchmarks; notably, the ccECP achieves the most favorable performance in the WMAD metric (Fig. \ref{fig:WMAD_in_elements}) while maintaining accuracy comparable to best legacy pseudopotentials in the MAD and LMAD assessments (Figs. \ref{fig:MAD_in_elements}--\ref{fig:LMAD_in_elements}).

A defining characteristic of Hf is the lanthanide contraction, where poor shielding of the $4f$ shell leads to an unexpectedly compact atomic radius\cite{jiang2010}. Although density of states and orbital energy analyses confirm that $4f$ states reside at lower semi-core energies \cite{spohn2009,jiang2010}, capturing their influence entirely through a 60-core pseudopotential has historically proven difficult \cite{koseki2002, balasubramanian1991}. Legacy ECPs often encountered challenges to accurately represent this core-valence split, resulting in inaccurate interactions with ligand valence orbitals in oxide systems \cite{minenkov2017, jaxon2025}. In order to circumvent these limitations, recent studies have defaulted to treating the $4f$ shell explicitly through computationally demanding AE frameworks \cite{romeu2025, bauschlicher2022}. 

\begin{figure*}[!htbp]
\centering
\begin{subfigure}{0.5\textwidth}
    \includegraphics[width=\textwidth]{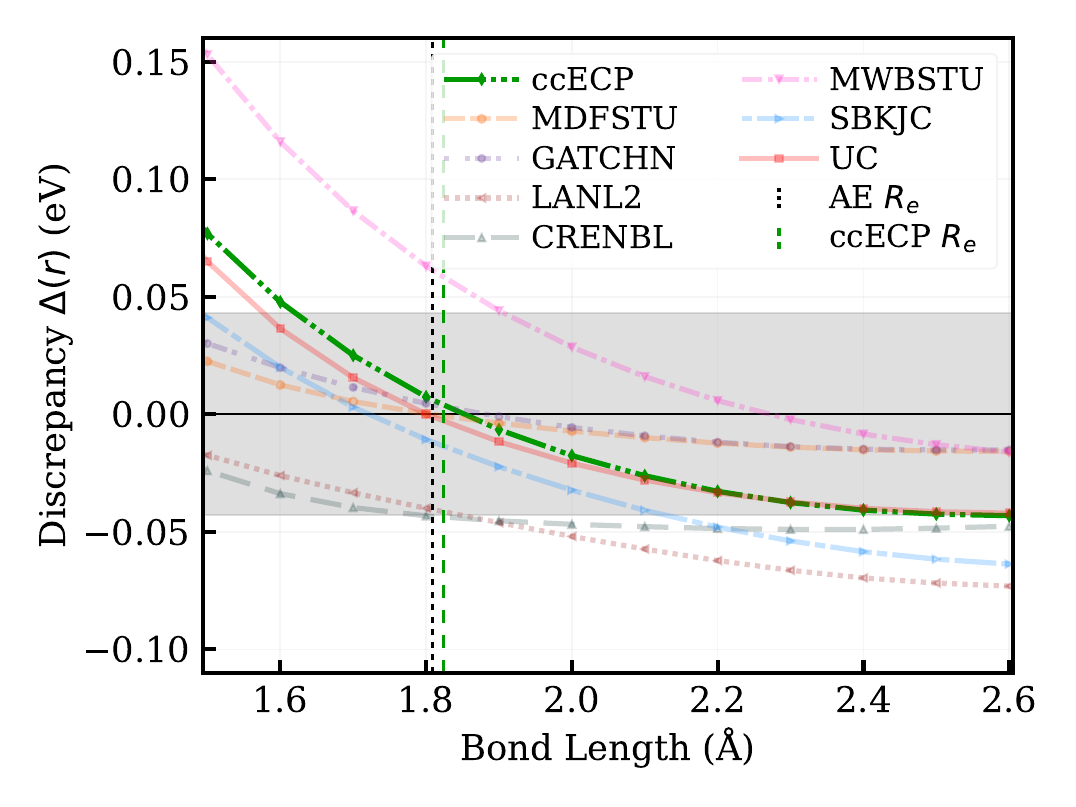}
    \caption{CCSD(T) binding energy discrepancies for HfH molecule}
    \label{fig:HfH}
\end{subfigure}\hfill
\begin{subfigure}{0.5\textwidth}
    \includegraphics[width=\textwidth]{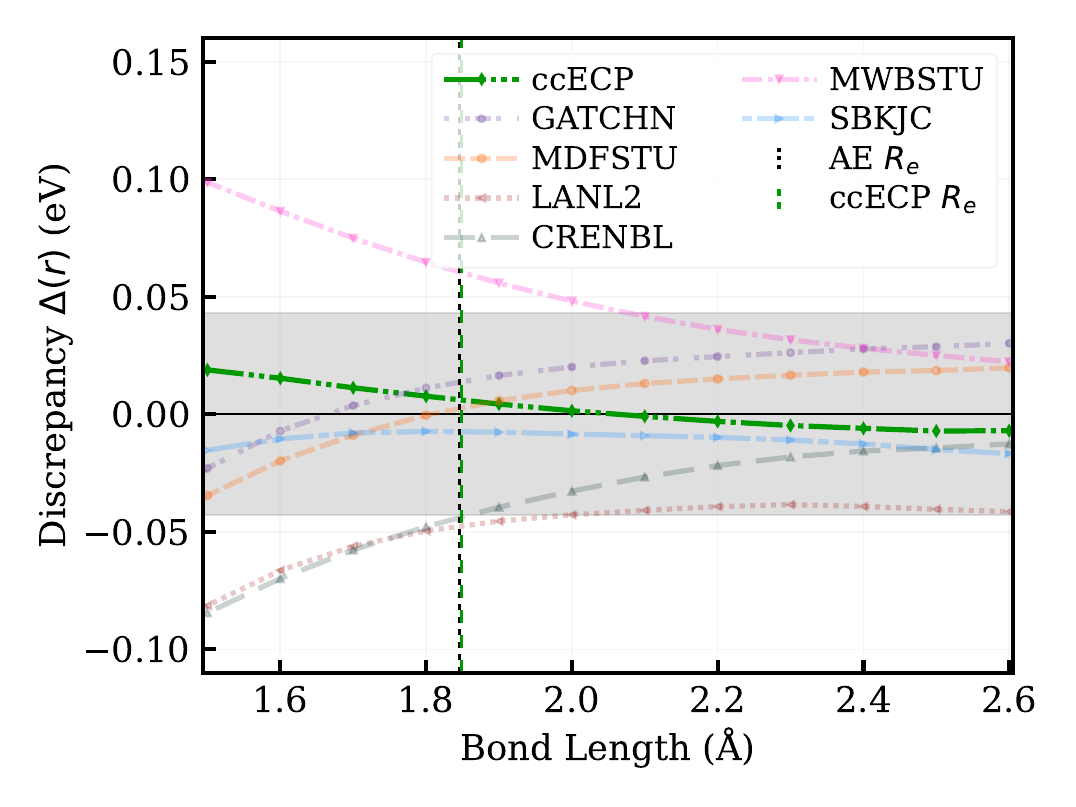}
    \caption{MRCI binding energy discrepancies for HfH molecule}
    \label{fig:HfH_mrci}
\end{subfigure}
\vspace{\baselineskip} 
\begin{subfigure}{0.5\textwidth}
    \includegraphics[width=\textwidth]{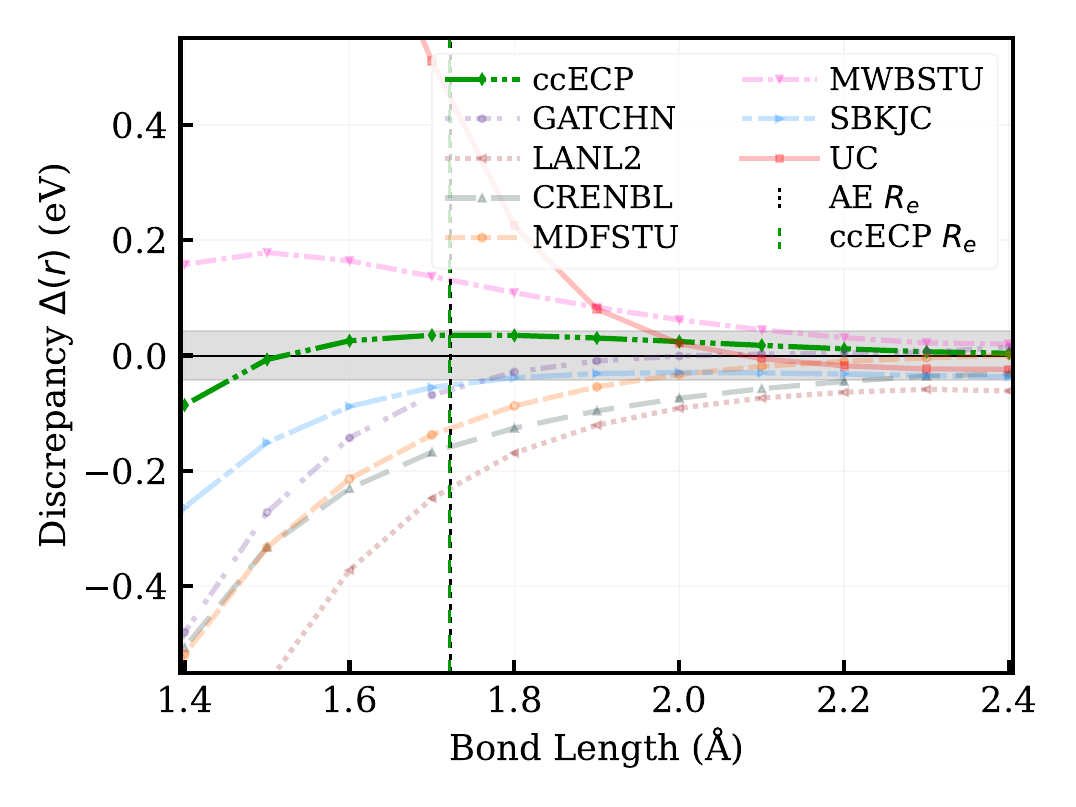}
    \caption{CCSD(T) binding energy discrepancies for HfO molecule}
    \label{fig:HfO}
\end{subfigure}
\caption{Binding energy discrepancies relative to the relativistic AE reference for (a) HfH and (b) HfO molecules. Markers denote the calculated discrepancies at discrete bond lengths, connected by lines to guide the eye, and the gray-shaded region illustrates the threshold for chemical accuracy. The vertical dashed lines mark the equilibrium bond length ($R_e$) extracted from the AE reference and ccECP. To evaluate relative performance, the ccECP is compared against available legacy ECPs that employ an equivalent 60-electron [[Kr]$4d^{10}4f^{14}$] core.}
\label{fig:Hf_mols}
\end{figure*}

Our molecular benchmarks demonstrate that these discrepancies are not merely limitations of core-valence separation, but are fundamentally addressed by shifting to a many-body formalism that explicitly captures core-core and core-valence correlations. This is most evident in HfO system in Fig. \ref{fig:HfO}, where the ccECP exhibits exceptional transferability, successfully bridging the gap between the highly polarized, multiple-bond interactions of the oxide and the less ionic, $s$-$d$ hybridized bonding of the hydride, a known challenge where competing static correlation effects often compromise performance \cite{ariyarathna2022}.  The ccECP achieves chemical accuracy ($<43$~meV) starting at $\sim$1.44 \AA~and maintains this fidelity through the equilibrium region and the dissociation limit. The resulting spectroscopic constants ($R_e = 1.722(2)$ \AA) are in excellent agreement with experimental values ($R_e \approx 1.723$ \AA\cite{bhartiya1986}) and recent AE benchmarks \cite{romeu2025}. 

For the HfH molecule, AE single-reference CCSD(T) calculations yield an equilibrium bond length ($R_e = 1.809(3)$~\AA) that notably underestimates the currently available experimental value currently available of $1.831$~\AA\cite{ram1994}. Under these same single-reference constraints, the ccECP result of  ($1.824(3)$~\AA) appears to align more closely with experiment; however, the shift in the AE baseline suggest that the single-configuration formalism may not fully account for the complex electronic manifold of HfH. Such effects are less pronounced in HfO, where the closed shell nature and the polarized valence density stabilizes a dominant configuration allowing CCSD(T) to maintain high accuracy for the oxide\cite{bauschlicher2022}.
To address this, previous studies have traditionally relied on multireference frameworks, such as multireference configuration interaction (MRCI) and second-order CI, to capture the substantial electron-degeneracy correlations inherent in HfH\cite{kher2024, koseki2002, balasubramanian1991}. Using MRCI, the AE $R_e$ shifts to $1.846(3)$~\AA, providing a baseline in better agreement with the experimental data.
The ccECP demonstrates excellent fidelity to this MRCI AE Hamiltonian, yielding an $R_e$ of $1.849(3)$~\AA. This close correspondence across both single- and multi-reference frameworks confirms that the ccECP accurately reproduces the underlying scalar relativistic many-body physics of hafnium while maintaining discrepancies well within the chemical accuracy window.

The ability to maintain a flat discrepancy curve across diverse bonding environments is critical for solid-state applications. In these $5d$ systems, strong lattice relaxations require the potential energy surface to be accurate also far from equilibrium, a requirement that standard mean-field approximations often fail to meet, leading to systematic discrepancies of the order of electronvolts \cite{chimata2019}. By providing a rigorously correlated description of many-body core interactions, the ccECP offers a robust foundation for the workflows necessary for high-fidelity predictive modeling. 
\subsubsection{Os}
For Osmium (Os, $Z_{\text{eff}}=16$), the ccECP treats 16 valence electrons ($5s^2 5p^6 5d^6 6s^2$). As shown in the atomic spectrum benchmarks (Figs. \ref{fig:MAD_in_elements}--\ref{fig:WMAD_in_elements}), the potential demonstrates a marked improvement over several legacy ECPs in the MAD metric, while providing on-par agreement with the best-performing legacy alternatives in the LMAD and WMAD assessments.

Osmium is characterized by an exceptionally dense landscape of low-lying electronic states, where the interplay of valence correlation and relativistic effects dictates the preferred bonding configurations\cite{yan2023,wang2026}. In OsO, the relativistic contraction of the 6s orbital and expansion of the 5d shell induce a molecular orbital reordering that distinguishes it from its lighter counterparts, FeO and RuO\cite{wang2026}. This relativistic regime is further evidenced by a unique bond length trend in which OsO is significantly more contracted than RuO)\cite{Omar2025}.
By retaining the $5s$ and $5p$ shells in the active space, the 60-core ccECP preserves the vital subvalence interaction and handles the high polarizability of the semi-core. 

\begin{figure*}[!htbp]
\centering
\begin{subfigure}{0.5\textwidth}
\includegraphics[width=\textwidth]{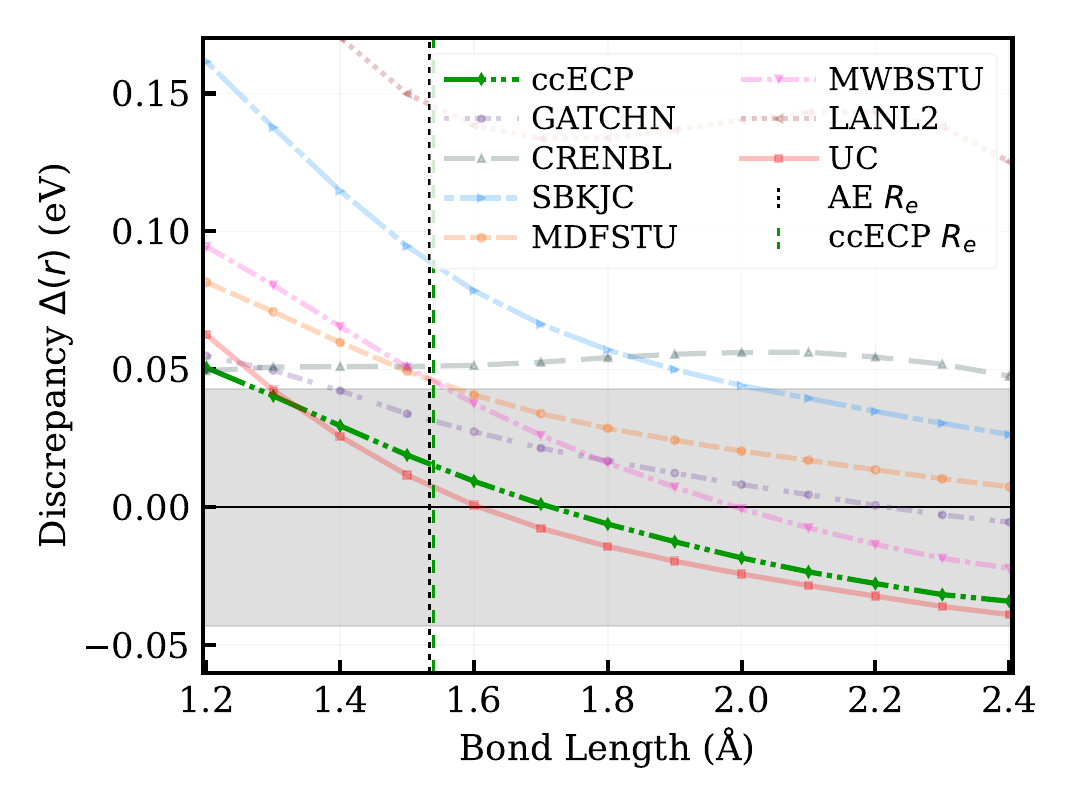}
\caption{CCSD(T) binding energy discrepancies for OsH molecule}
\label{fig:OsH}
\end{subfigure}%
\begin{subfigure}{0.5\textwidth}
\includegraphics[width=\textwidth]{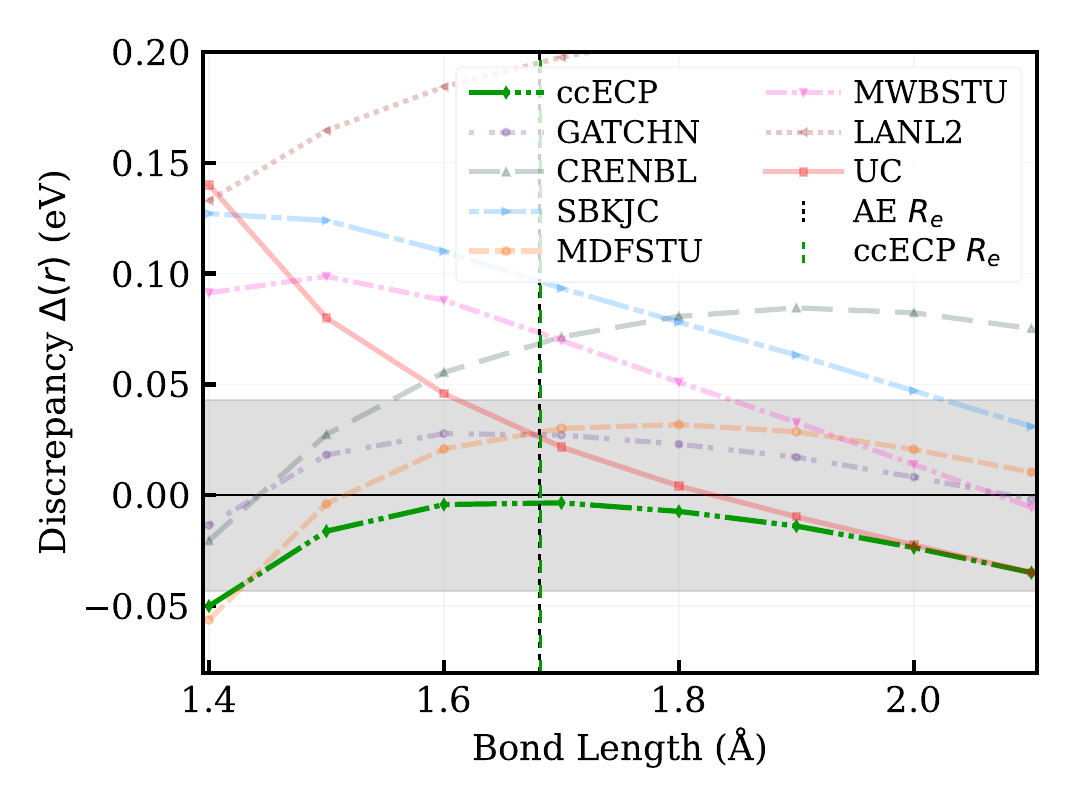}
\caption{CCSD(T) binding energy discrepancies for OsO molecule}
\label{fig:OsO}
\end{subfigure}
\caption{Binding energy discrepancies relative to the relativistic AE reference for (a) OsH and (b) OsO molecules. The graphical conventions are identical to those detailed in Fig. \ref{fig:Hf_mols}.
The ccECP is compared against available legacy ECPs that employ an equivalent 60-electron [[Kr]$4d^{10}4f^{14}$] core.}
\label{fig:Os_mols}
\end{figure*}

The robustness of this approach is most evident in the hydride (OsH) benchmarks (Fig. \ref{fig:OsH}), where the ccECP is the only potential to perform within chemical accuracy ($<43$~meV discrepancy) throughout the entire potential energy surface, effectively mitigating the significant underbinding at compressed bond lengths observed in legacy ECPs. Although the AE frozen-core UC approximation may yield acceptable results for the hydride, its transferability is compromised when tested on the mono-oxide (OsO, Fig. \ref{fig:OsO}), exhibiting significant deviations at compressed bond lengths. Of course, this is of significant consideration for high pressure solids and signals that in all-electron studies the outer core states have to be treated in a correlated framework as well. In contrast, the ccECP recovers the behavior of the AE Hamiltonian with remarkable precision, maintaining discrepancies below 1 kcal/mol across the dissociation curve. Furthermore, the equilibrium bond length of the ccECP ($1.680(9)$~\AA) is effectively indistinguishable from the AE scalar-relativistic CCSD(T) ($1.681(8)$~\AA), with both results aligning with the experimental value of 1.677~\AA\cite{wang2026}.

This structural accuracy provides a critical advantage for high-level many-body calculations. For heavy transition metal oxides, intricate $d$-$p$ hybridization and nodal structures strongly prescribe correlation physics \cite{wagner2007}. The smooth density representation inherent to the ccECP formalism is essential here; by replacing the deep core shells whose large energy contributions and orbital derivatives introduce severe scaling statistical fluctuations \cite{foulkes2001}; the ccECP significantly alleviates both the biases and the noise that impact QMC methods. Consequently, it provides the valence representation necessary to properly capture complex many-body effects and enables predictive modeling of heavy-element systems without the prohibitive computational cost of AE treatments.
\subsubsection{Hg}
\begin{figure*}[!htbp]
\centering
\begin{subfigure}{0.5\textwidth}
\includegraphics[width=\textwidth]{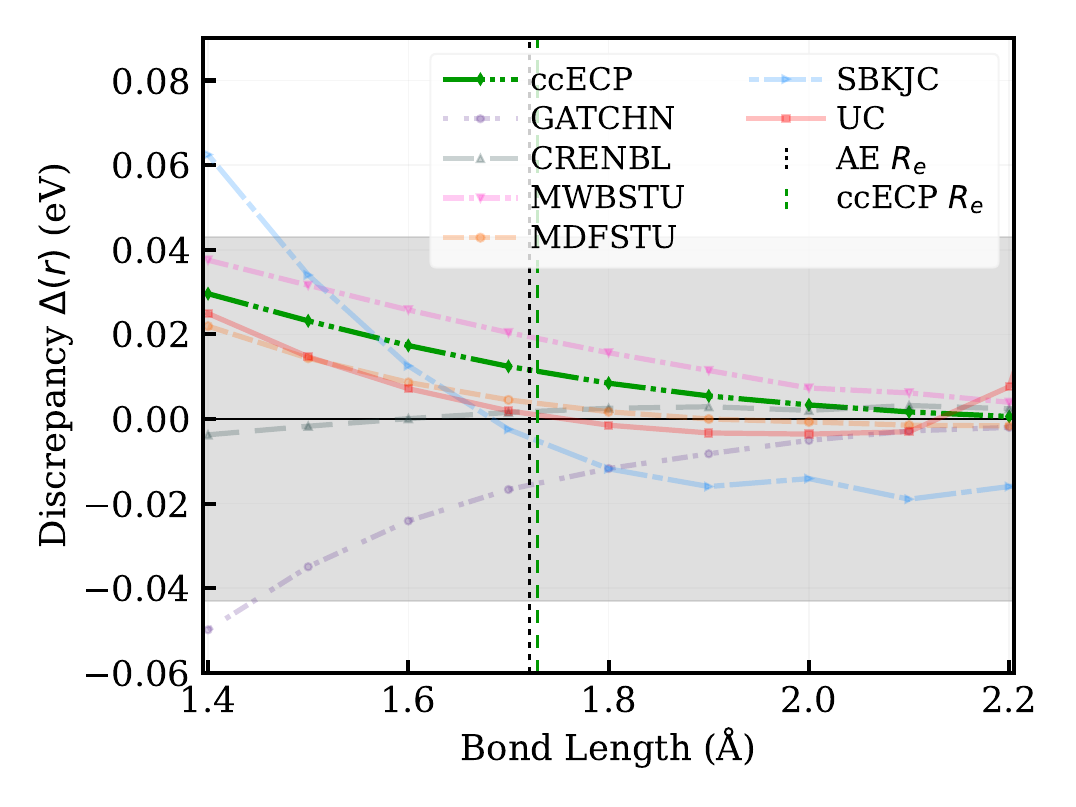 }
\caption{CCSD(T) binding energy discrepancies for HgH molecule}
\label{fig:HgH_20}
\end{subfigure}%
\begin{subfigure}{0.5\textwidth}
\includegraphics[width=\textwidth]{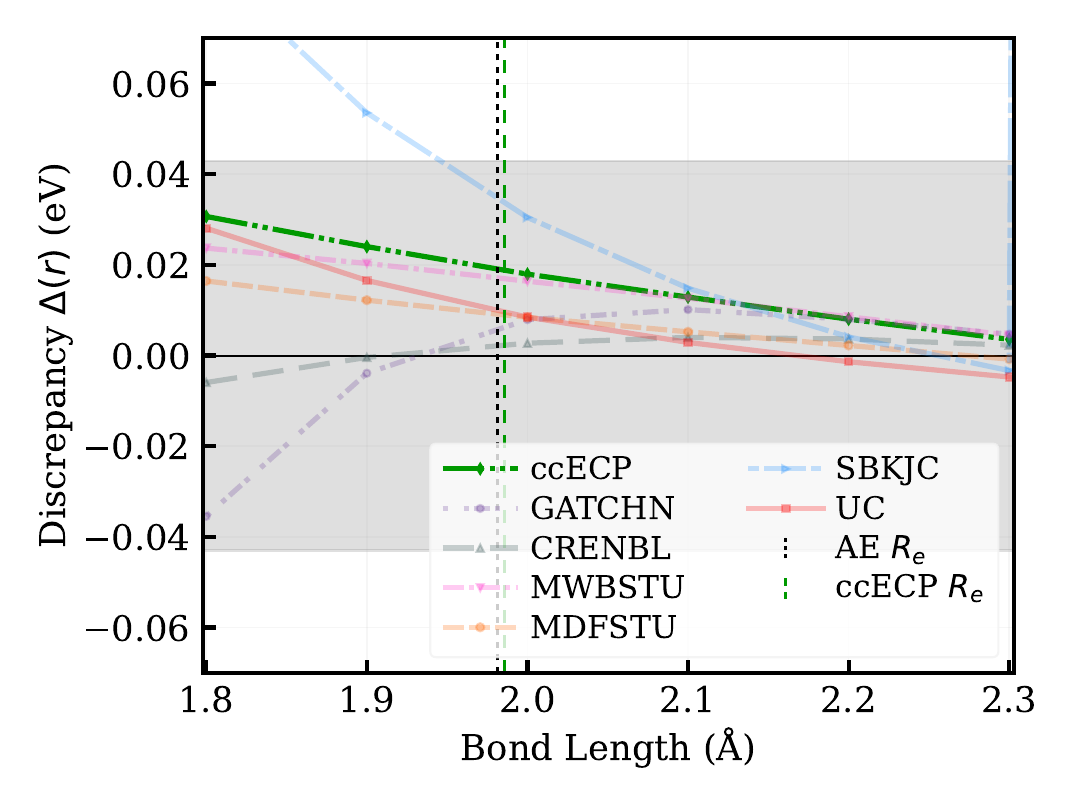}
\caption{CCSD(T) binding energy discrepancies for HgO$^{+}$ molecular ion}
\label{fig:HgO_20}
\end{subfigure}
\caption{Binding energy discrepancies relative to the relativistic AE reference for (a) HgH and (b) HgO$^{+}$ molecules. The graphical conventions are identical to those detailed in Fig. \ref{fig:Hf_mols}.
The ccECP is compared against available legacy ECPs that employ an equivalent 60-electron [[Kr]$4d^{10}4f^{14}$] core.}
\label{fig:Hg_20_mols}
\end{figure*}

Consistent with the strategy applied to heavy $5d$ transition metals, small-core mercury (Hg) ccECP leaves 20 valence electrons ($5s^{2}5p^{6}5d^{10}6s^{2}$) outside the [[Kr]$4d^{10}4f^{14}$] core. Given this expansive valence space, the discrepancy margins for most small-core approximations are naturally minimized, yielding high accuracy across atomic and molecular benchmarks. The ccECP matches this high performance in the atomic spectrum (Figs. \ref{fig:MAD_in_elements}-\ref{fig:WMAD_in_elements}) and maintains negligible discrepancies in the binding curves of the weakly bound HgH and the highly polarized HgO$^{+}$ (Fig. \ref{fig:Hg_20_mols}). This 60-core variant is particularly critical for states that require deep ionization past the third ionization potential, where the $5d$ shell is explicitly broken.

\begin{figure*}[!htbp]
\centering
\begin{subfigure}{0.5\textwidth}
\includegraphics[width=\textwidth]{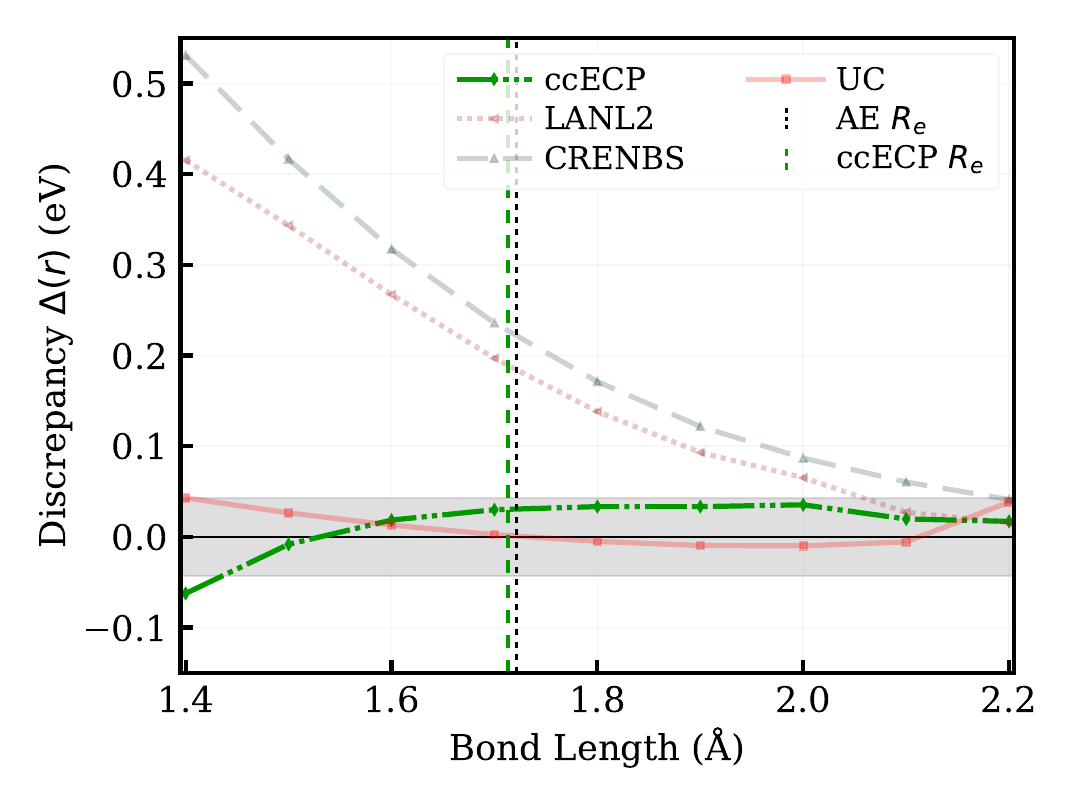 }
\caption{CCSD(T) binding energy discrepancies for HgH molecule}
\label{fig:HgH_12}
\end{subfigure}%
\begin{subfigure}{0.5\textwidth}
\includegraphics[width=\textwidth]{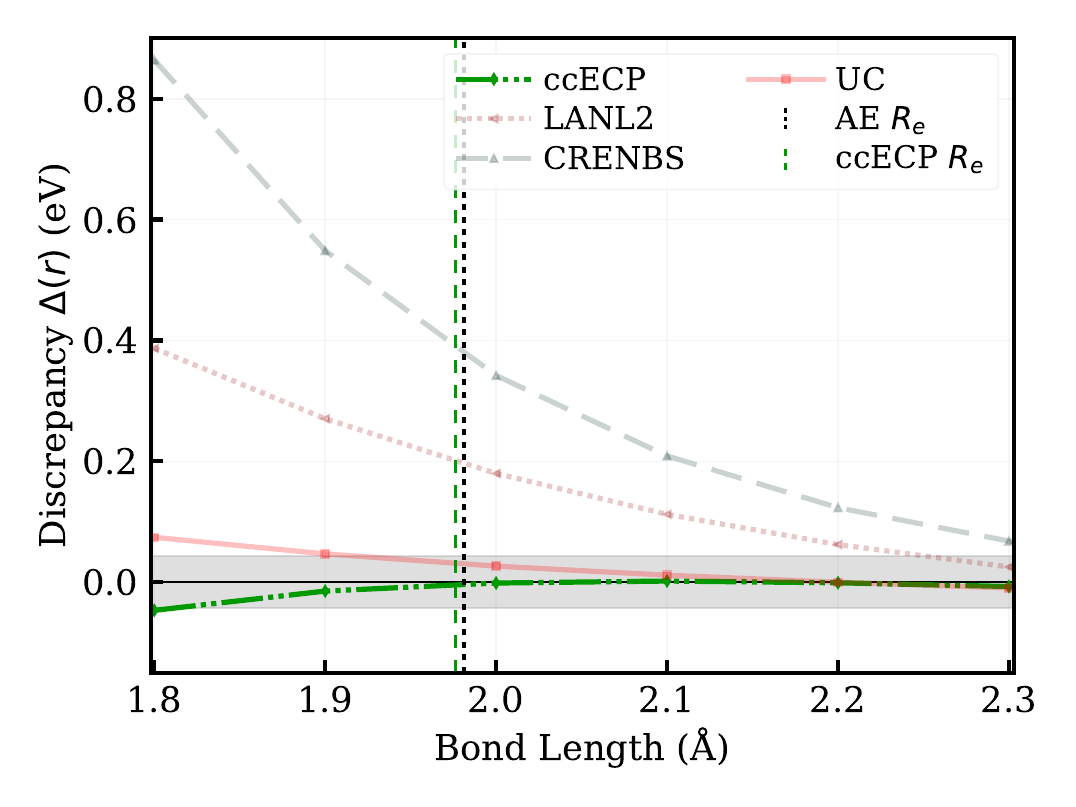}
\caption{CCSD(T) binding energy discrepancies for HgO$^{+}$ molecular ion}
\label{fig:HgO_12}
\end{subfigure}
\caption{Binding energy discrepancies relative to the relativistic AE reference for (a) HgH and (b) HgO$^{+}$ molecules. The graphical conventions are identical to those detailed in Fig. \ref{fig:Hf_mols}.
The ccECP is compared against available legacy ECPs that employ an equivalent 68-electron [[Xe]4$f^{14}$] core.}
\label{fig:Hg_12_mols}
\end{figure*}

To optimally balance accuracy with computational efficiency, we also developed a medium-core (68-core) variant that leaves 12 valence electrons ($5d^{10}6s^{2}$) outside a [[Xe]$4f^{14}$] core. Theoretical studies of mercury explicitly highlight that $5d$ correlation is a fundamental requirement for bonding; without its inclusion, the system intrinsically fails to bind\cite{gaston2006}. Furthermore, the 68-core partition has long been recognized as the rigorous theoretical standard required to capture the relatively diffuse nature of the Hg $5d$ shell \cite{dolg1991}. Although identified as the optimal partition for efficient molecular calculations, accurately capturing the explicit correlation of this valence space has historically proven difficult \cite{mosyagin2005}. In our tests, the 68-core ccECP successfully mirrors the small-core's chemical accuracy across the MAD metrics (Figs. \ref{fig:MAD_in_elements}-\ref{fig:WMAD_in_elements}) and the molecular HgH and HgO$^+$ dissociation curves (Fig. \ref{fig:Hg_12_mols}), serving as the most practical choice for systems that do not deeply ionize the $5d$ shell.

In contrast, legacy 68-core pseudopotentials struggle to manage this sensitive core-valence boundary. Previous analyses demonstrate that legacy 12-electron pseudo-approximations cannot satisfactorily describe core-valence exchange interactions because smoothed valence functions exhibit incorrect behavior in the outer-core region \cite{mosyagin2000, mosyagin2005}. Consequently, legacy ECPs suffer from severe underbinding at compressed geometries, straying  from chemical accuracy even at the equilibrium bond length. This aligns with historical benchmarks that demonstrate that legacy ECPs systematically underestimate the dissociation energy of weakly bound systems such as HgH \cite{dolg1991}. Although the UC potential performs adequately near equilibrium for HgH, it exhibits discrepancies greater than 1 kcal/mol at the compressed bond lengths of HgO$^+$. By explicitly incorporating constraints to enforce Coulomb cusp cancellation at the origin, the ccECPs successfully regularize these interactions, resolving the physical shortcomings of legacy pseudopotentials.
\begin{table*}
\small
\centering
\caption{
ccECPs SOREP parameters for selected $5d$ elements.
$Z_{\text{eff}}$ denotes the effective core charge.
The local channel $L$ corresponds to the highest angular momentum, while the non-local channels are defined for $\ell<L$. 
Gaussian parameters $\alpha_{\ell k}$ and $\beta_{\ell k}$ follow the functional forms defined in Section \ref{Methods}.}
\label{tab:selected_5d_params}
\begin{tabular}{cccccrrccccccrrr}
\hline
\hline
\multicolumn{1}{c}{Atom} & \multicolumn{1}{c}{$Z_{\rm eff}$} & \multicolumn{1}{c}{Hamiltonian} & \multicolumn{1}{c}{$\ell$} & \multicolumn{1}{c}{$n_{\ell k}$} & \multicolumn{1}{c}{$\alpha_{\ell k}$} & \multicolumn{1}{c}{$\beta_{\ell k}$} & & \multicolumn{1}{c}{Atom} & \multicolumn{1}{c}{$Z_{\rm eff}$} & \multicolumn{1}{c}{Hamiltonian} & \multicolumn{1}{c}{$\ell$} & \multicolumn{1}{c}{$n_{\ell k}$} & \multicolumn{1}{c}{$\alpha_{\ell k}$} & \multicolumn{1}{c}{$\beta_{\ell k}$} \\
\hline
   &    &      &   &   &             &              &&    &    &      &   &   &             &              \\
Hf & 11 & AREP & 0 & 2 &    7.926751 &  178.228590  && Os & 16 & AREP & 0 & 2 &   11.500246 &  471.040730  \\
   &    &      & 0 & 2 &    1.832226 &   -1.238327  &&    &    &      & 0 & 2 &    3.485240 &   17.813049  \\
   &    &      & 1 & 2 &    9.591748 &  218.375370  &&    &    &      & 1 & 2 &    9.803703 &  265.267755  \\
   &    &      & 1 & 2 &    3.385934 &    3.253183  &&    &    &      & 1 & 2 &    4.648167 &   48.510069  \\
   &    &      & 1 & 4 &    9.593456 &  401.886856  &&    &    &      & 2 & 2 &    6.202758 &  107.920976  \\
   &    &      & 2 & 2 &    6.172055 &  106.940492  &&    &    &      & 2 & 2 &    4.055057 &   31.437960  \\
   &    &      & 2 & 2 &   13.764073 &  -48.529458  &&    &    &      & 3 & 2 &    2.502922 &   16.905644  \\
   &    &      & 2 & 4 &    6.178253 &   83.536935  &&    &    &      & 3 & 2 &    4.013976 &   17.858142  \\
   &    &      & 3 & 2 &    1.917753 &   12.670555  &&    &    &      & 4 & 1 &   13.922849 &   16.000000  \\
   &    &      & 3 & 2 &    3.199668 &   14.014442  &&    &    &      & 4 & 3 &   14.540135 &  222.765582  \\
   &    &      & 4 & 1 &    6.602477 &   12.000000  &&    &    &      & 4 & 2 &    3.498721 &   -8.886788  \\
   &    &      & 4 & 3 &    9.156737 &   79.229729  &&    &    &      & 4 & 2 &    4.138579 &  -10.831331  \\
   &    &      & 4 & 2 &    7.992198 &   -0.262403  &&    &    &      &   &   &             &              \\
   &    &      & 4 & 2 &    2.356352 &   -2.154885  &&    &    &      &   &   &             &              \\
   &    &      &   &   &             &              &&    &    &      &   &   &             &              \\
   &    & SO   & 1 & 2 &    7.348524 & -202.108107  &&    &    & SO   & 1 & 2 &   10.560297 & -176.558736  \\
   &    &      & 1 & 2 &    9.040090 &  212.179008  &&    &    &      & 1 & 2 &    9.921793 &  176.477661  \\
   &    &      & 1 & 4 &    9.378543 &   -0.248195  &&    &    &      & 1 & 2 &    6.631575 &  -23.425209  \\
   &    &      & 1 & 4 &    8.245007 &    0.444413  &&    &    &      & 1 & 2 &    4.940591 &   22.924627  \\
   &    &      & 1 & 2 &    1.277345 &    3.874243  &&    &    &      & 2 & 2 &    7.024418 &  -44.783589  \\
   &    &      & 1 & 2 &    2.365402 &    0.652432  &&    &    &      & 2 & 2 &    6.736237 &   44.765213  \\
   &    &      & 2 & 2 &    5.361294 &  -43.052898  &&    &    &      & 2 & 2 &    3.812040 &   -5.561585  \\
   &    &      & 2 & 2 &    5.506767 &   43.998279  &&    &    &      & 2 & 2 &    3.627003 &    5.486400  \\
   &    &      & 2 & 4 &    5.936054 &    0.014699  &&    &    &      & 3 & 2 &    2.877899 &   -5.543503  \\
   &    &      & 2 & 4 &    5.709982 &    0.057848  &&    &    &      & 3 & 2 &    2.723639 &    5.561293  \\
   &    &      & 2 & 2 &    0.539404 &    0.097633  &&    &    &      &   &   &             &              \\
   &    &      & 2 & 2 &    0.484937 &   -0.031629  &&    &    &      &   &   &             &              \\
   &    &      & 3 & 2 &    3.508029 &   -2.893197  &&    &    &      &   &   &             &              \\
   &    &      & 3 & 2 &    2.378979 &    3.130280  &&    &    &      &   &   &             &              \\
   &    &      &   &   &             &              &&    &    &      &   &   &             &              \\
   &    &      &   &   &             &              &&    &    &      &   &   &             &              \\
Hg & 20 & AREP & 0 & 2 &   12.953229 &  275.737658  && Hg & 12  &AREP & 0 & 2 &    3.940278 &   62.653492  \\
   &    &      & 0 & 2 &    6.450052 &   49.091794  &&    &    &      & 0 & 2 &    1.621953 &   38.099901  \\
   &    &      & 1 & 2 &   10.423007 &  241.541717  &&    &    &      & 1 & 2 &    2.717680 &   55.807042  \\
   &    &      & 1 & 2 &    5.386795 &   27.395576  &&    &    &      & 1 & 2 &    1.050882 &   17.020907  \\
   &    &      & 2 & 2 &    7.933648 &  127.872345  &&    &    &      & 2 & 2 &    7.019022 &   69.601558  \\
   &    &      & 2 & 2 &    4.184795 &   16.615955  &&    &    &      & 2 & 2 &    2.084342 &   -4.950778  \\
   &    &      & 3 & 2 &    4.085791 &   10.182498  &&    &    &      & 3 & 2 &    6.195706 &   31.726110  \\
   &    &      & 3 & 2 &    3.685791 &   20.182498  &&    &    &      & 3 & 2 &    0.623552 &   -0.333353  \\
   &    &      & 4 & 1 &   10.186067 &   20.000000  &&    &    &      & 4 & 1 &    3.007525 &   12.000000  \\
   &    &      & 4 & 3 &   10.591685 &  203.721339  &&    &    &      & 4 & 3 &    2.986195 &   36.090300  \\
   &    &      & 4 & 2 &   10.058090 & -123.870123  &&    &    &      & 4 & 2 &    2.284716 &   -6.965301  \\
   &    &      & 4 & 2 &   10.591786 &   -7.744438  &&    &    &      & 4 & 2 &    1.792313 &  -14.937295  \\
   &    &      &   &   &             &              &&    &    &      &   &   &             &              \\
   &    & SO   & 1 & 2 &   11.318097 & -161.013792  &&    &    & SO   & 1 & 2 &    0.990798 &   -5.218825  \\
   &    &      & 1 & 2 &   10.199506 &  161.035062  &&    &    &      & 1 & 2 &    0.849331 &    5.558408  \\
   &    &      & 1 & 2 &    5.946689 &  -18.165839  &&    &    &      & 1 & 2 &    0.466015 &    0.421688  \\
   &    &      & 1 & 2 &    5.009057 &   18.369198  &&    &    &      & 1 & 2 &    0.329840 &   -0.108995  \\
   &    &      & 2 & 2 &    8.405671 &  -51.137372  &&    &    &      & 2 & 2 &    0.852968 &   -2.901239  \\
   &    &      & 2 & 2 &    8.216516 &   51.138181  &&    &    &      & 2 & 2 &    0.526634 &    2.938491  \\
   &    &      & 2 & 2 &    4.009298 &   -6.562697  &&    &    &      & 3 & 2 &    0.558669 &    0.964790  \\
   &    &      & 2 & 2 &    3.799262 &    6.544385  &&    &    &      & 3 & 2 &    0.806490 &   -1.739464  \\
   &    &      & 3 & 2 &    3.269780 &   -6.286626  &&    &    &      &   &   &             &              \\
   &    &      & 3 & 2 &    3.211865 &    6.246768  &&    &    &      &   &   &             &              \\
   &    &      &   &   &             &              &&    &    &      &   &   &             &              \\
\hline
\hline
\end{tabular}
\end{table*}
\subsection{Selected $6p$ elements}
For the $6p$ row elements, the ccECPs primarily utilize a [[Xe]$4f^{14}5d^{10}$] large-core definition, explicitly treating only the $6s$ and $6p$ valence electrons ($Z_{\text{eff}} = 3$--$8$). To rigorously evaluate the trade-offs between computational efficiency and the treatment of semi-core correlation at the onset of the row, we additionally constructed small- and medium-core variants for thallium (Tl), utilizing [[Kr]$4d^{10}4f^{14}$] ($Z_{\text{eff}} = 21$) and [[Xe]$4f^{14}$] ($Z_{\text{eff}} = 13$) cores, respectively. Across the series, these ccECPs demonstrate exceptional accuracy in both atomic spectra and molecular environments, effectively capturing the energetic consequences of profound relativistic shielding and spin-orbit coupling inherent to heavy main-group elements.
\subsubsection{Tl}
To provide optimal flexibility for varying computational demands, we developed three ccECP variants for Thallium. All three variants demonstrate exceptional accuracy against relativistic AE CCSD(T) benchmarks, with the choice of pseudopotential depending on the specific physical demands of the target application.

\begin{figure*}[!htbp]
\centering
\begin{subfigure}{0.5\textwidth}
\includegraphics[width=\textwidth]{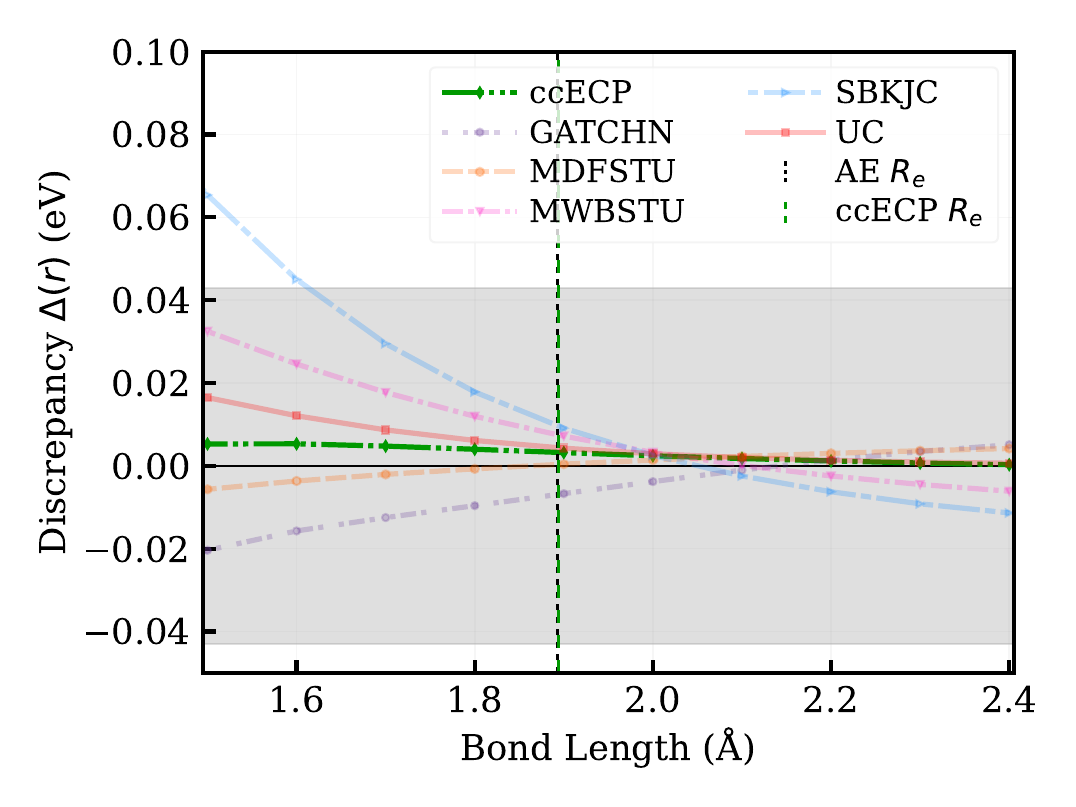}
\caption{CCSD(T) binding energy discrepancies for TlH molecule}
\label{fig:TlH_60}
\end{subfigure}%
\begin{subfigure}{0.5\textwidth}
\includegraphics[width=\textwidth]{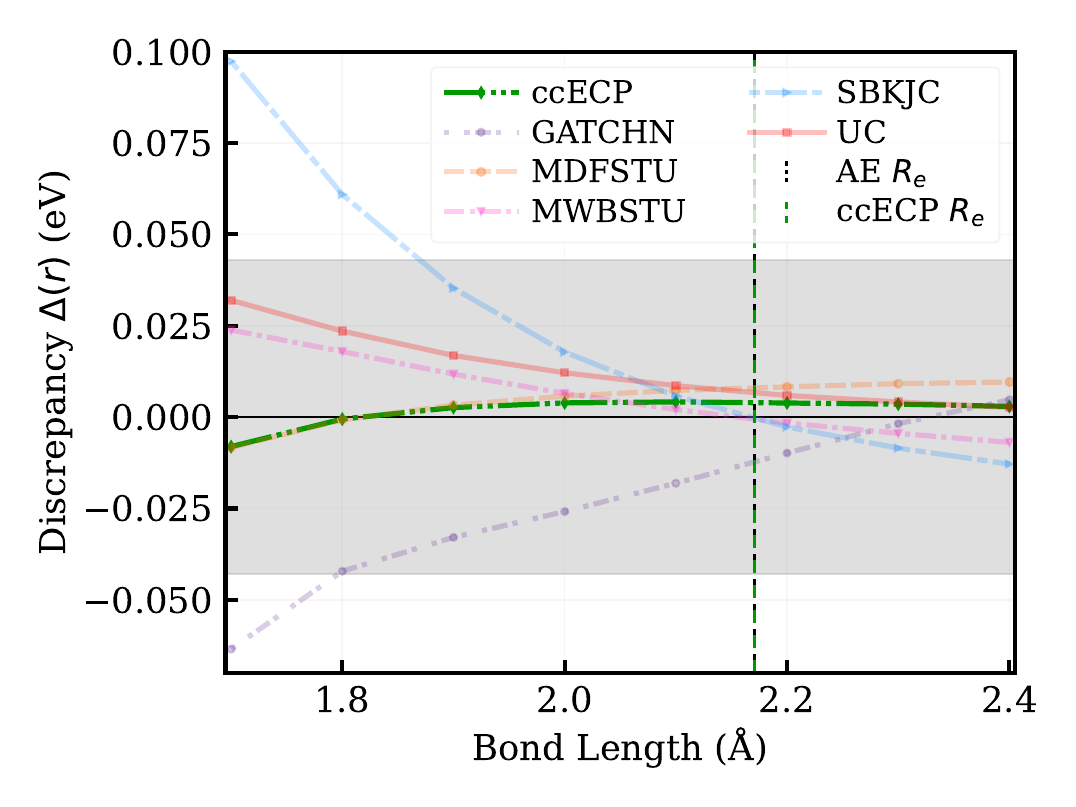}
\caption{CCSD(T) binding energy discrepancies for TlO molecule}
\label{fig:TlO_60}
\end{subfigure}
\caption{Binding energy discrepancies relative to the relativistic AE reference for (a) TlH and (b) TlO molecules. The graphical conventions are identical to those detailed in Fig. \ref{fig:Hf_mols}.
The ccECP is compared against available legacy ECPs that employ an equivalent 60-electron [[Kr]4$d^{10}$4$f^{14}$] core.}
\label{fig:Tl60_mols}
\end{figure*}

The small-core ccECP ($Z_{\text{eff}}=21$) explicitly treats the $5s, 5p, 5d, 6s,$ and $6p$ shells. Historically, researchers have often been compelled to utilize this expanded 21-electron space because legacy approximations of smaller valence spaces could not achieve good accuracy in capturing intricate core-valence and relativistic physics required for chemical precision \cite{titov2000}. Serving as a nearly exact reference that approaches the AE scalar relativistic limit \cite{tang2024}, The ccECP provides a robust alternative to the BFD formalism by rigorously regularizing the potential at the origin. Our benchmarks demonstrate that the 21-electron ccECP recovers the AE CCSD(T) equilibrium bond lengths with sub-picometer fidelity. It also achieves an atomic LMAD of $\sim$0.02~eV (Fig. \ref{fig:LMAD_in_elements}) and practically negligible discrepancy in the molecular benchmarks (Fig. \ref{fig:Tl60_mols}), establishing it as the definitive choice for benchmark studies and states requiring deep ionizations beyond the fourth ionization potential.

\begin{figure*}[!htbp]
\centering
\begin{subfigure}{0.5\textwidth}
\includegraphics[width=\textwidth]{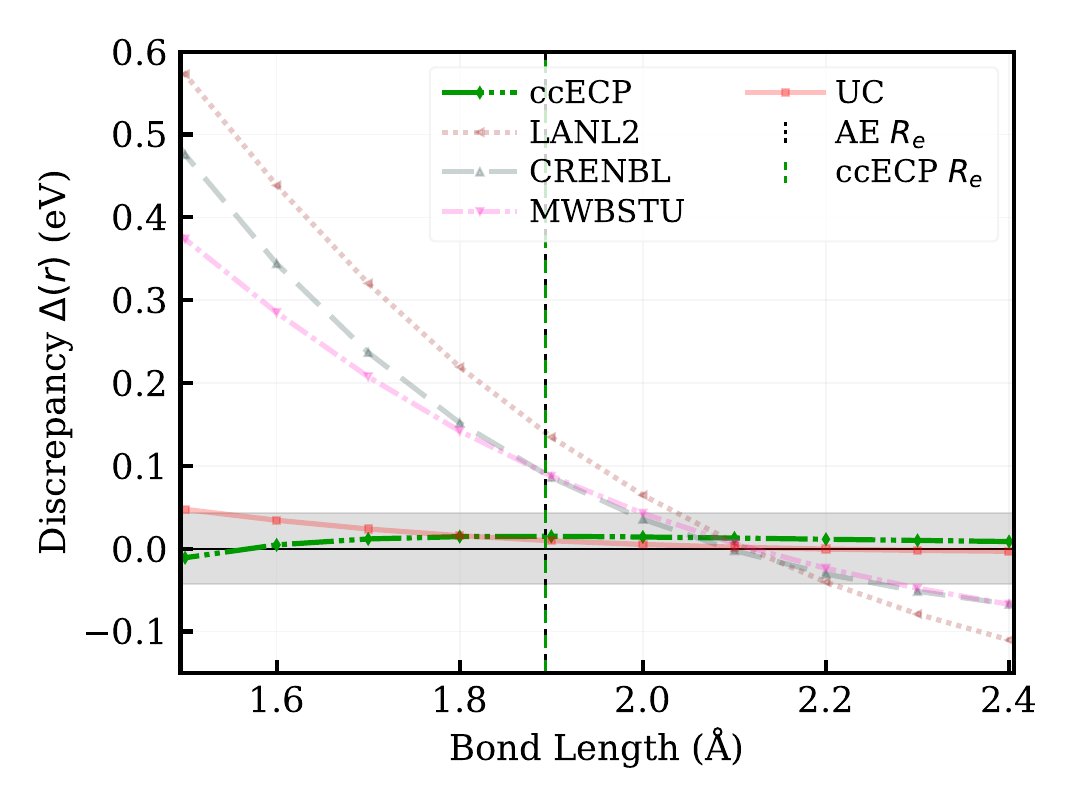}
\caption{CCSD(T) binding energy discrepancies for TlH molecule}
\label{fig:TlH_13}
\end{subfigure}%
\begin{subfigure}{0.5\textwidth}
\includegraphics[width=\textwidth]{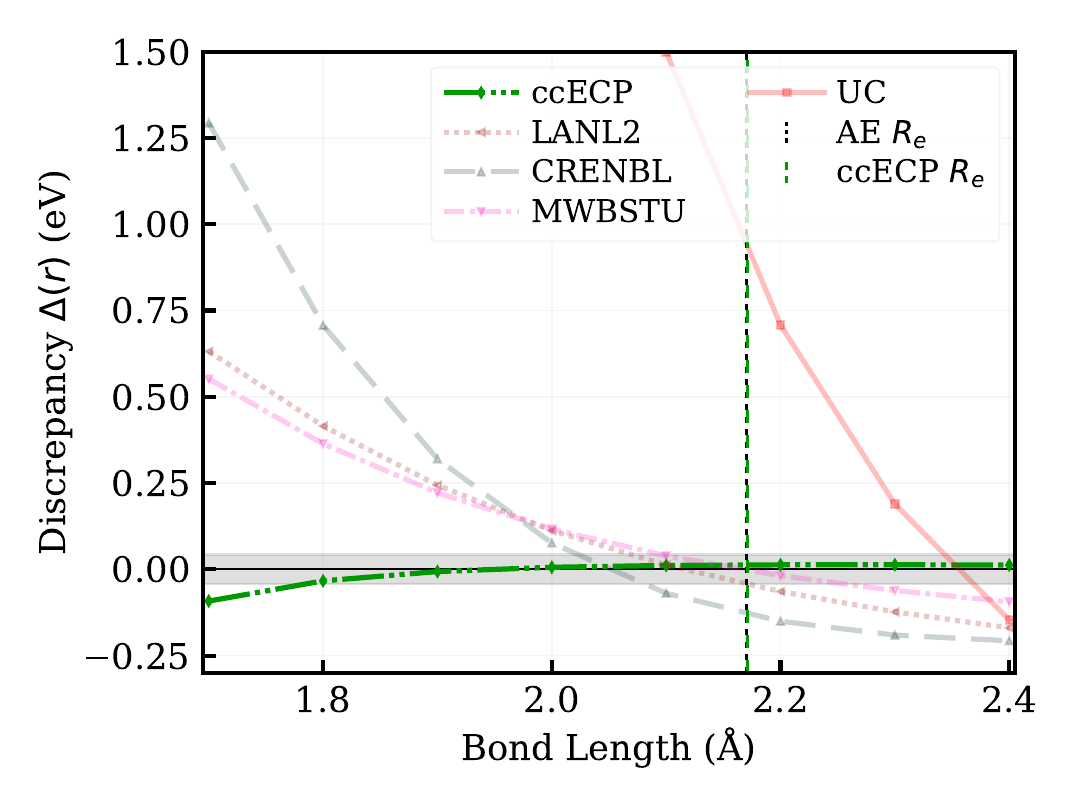}
\caption{CCSD(T) binding energy discrepancies for TlO molecule}
\label{fig:TlO_13}
\end{subfigure}
\caption{Binding energy discrepancies relative to the relativistic AE reference for (a) TlH and (b) TlO molecules. The graphical conventions are identical to those detailed in Fig. \ref{fig:Hf_mols}.
The ccECP is compared against available legacy ECPs that employ an equivalent 68-electron [[Xe]4$f^{14}$] core.
}
\label{fig:Tl13_mols}
\end{figure*}

To optimally balance accuracy with computational efficiency, we developed a medium-core variant ($Z_{\text{eff}}=13$, [[Xe]$4f^{14}$]) that explicitly correlates the $5d$ shell. Relativistic assessments identify this 13-electron partition as the theoretical optimum for most applications; $5d$ correlation typically accounts for roughly 10\% of the subvalence physics, whereas $5s$ and $5p$ contributions are significantly smaller and less active in chemical processes\cite{fleig2020}. Unfortunately, legacy 13-electron potentials have historically been limited to reach this accuracy limit \cite{titov2000}. Our benchmarks confirm this accuracy ceiling where legacy pseudopotential discrepancies $\Delta$(r)$\sim$0.1-0.2~eV at equilibrium for TlH (Fig. \ref{fig:TlH_13}) and exhibit severe underbinding of $\sim$0.5~eV at compressed bond lengths. For TlH, the AE and ccECP results align within 0.02~\AA\ of the experiment (1.870~\AA), with the remaining discrepancies likely due to spin-orbit induced bond shortening. The 13-electron ccECP consistently recovers this AE level and maintains a near-zero discrepancy across the potential energy surfaces of TlH and TlO. Notably, the frozen-core AE-UC method suffers a catastrophic breakdown in the highly polarized TlO system (Fig. \ref{fig:TlO_13}) ($>1.5$~eV discrepancy), proving the 68-core ccECP as the most practical and reliable choice for high-fidelity molecular applications.

\begin{figure*}[!htbp]
\centering
\begin{subfigure}{0.5\textwidth}
\includegraphics[width=\textwidth]{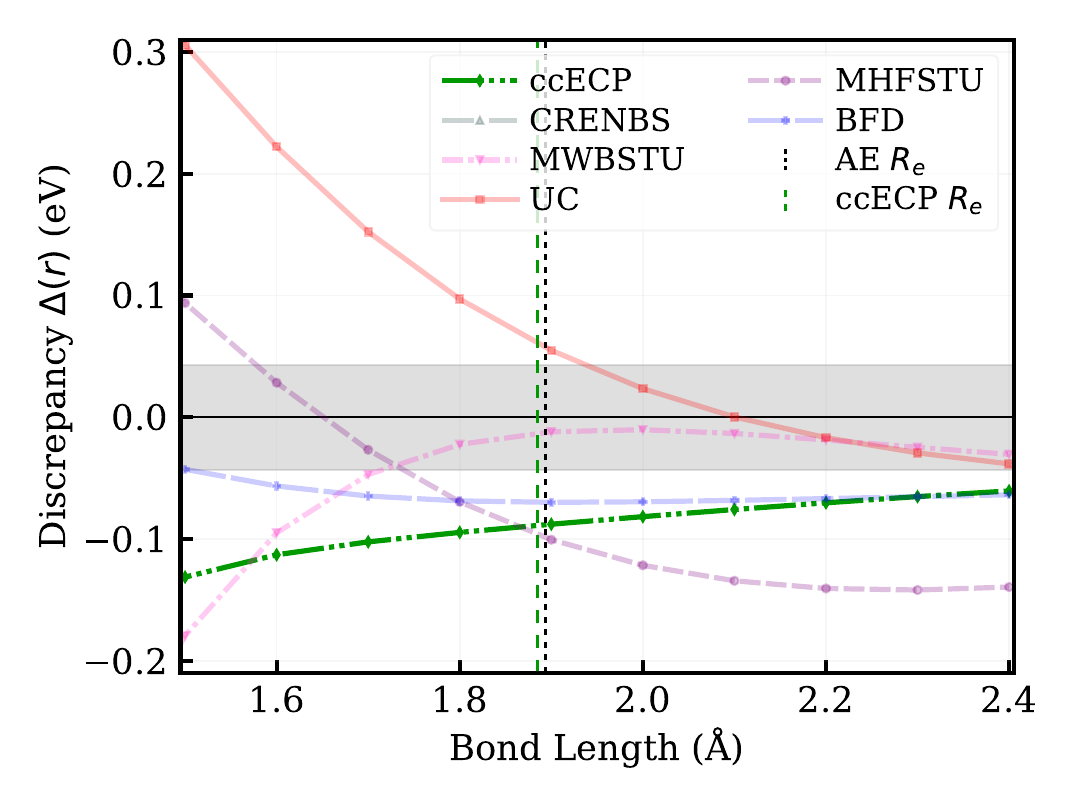}
\caption{CCSD(T) binding energy discrepancies for TlH molecule}
\label{fig:TlH_3}
\end{subfigure}%
\begin{subfigure}{0.5\textwidth}
\includegraphics[width=\textwidth]{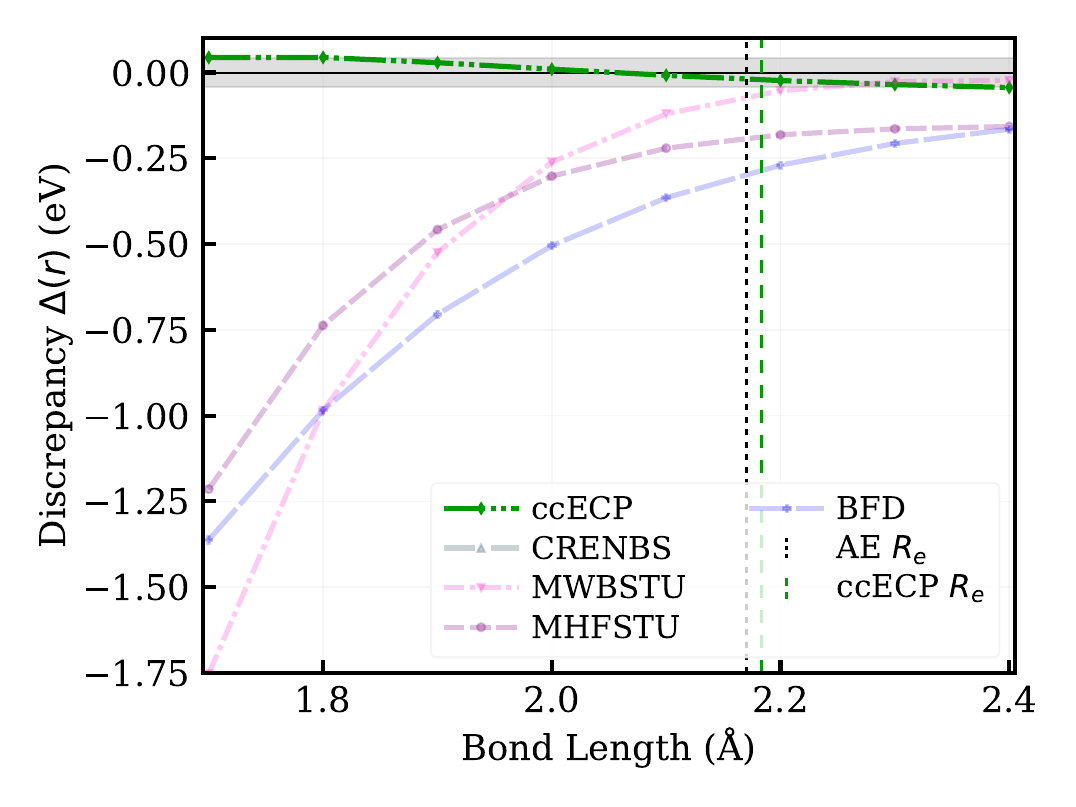}
\caption{CCSD(T) binding energy discrepancies for TlO molecule}
\label{fig:TlO_3}
\end{subfigure}
\caption{Binding energy discrepancies relative to the relativistic AE reference for (a) TlH and (b) TlO molecules. The graphical conventions are identical to those detailed in Fig. \ref{fig:Hf_mols}.
The ccECP is compared against available legacy ECPs that employ an equivalent 78-electron [[Xe]4$f^{14}$5$d^{10}$] core.
}
\label{fig:Tl3_mols}
\end{figure*}

The  challenge of the large-core variant ($Z_{\text{eff}}=3$, [[Xe]$4f^{14}5d^{10}$]) lies in capturing the $6s^{2}6p^{1}$ valence space while the highly polarizable $5d$ shell remains frozen. Absorbing the $5d$ shell intrinsically limits the electronic polarization required to fully resolve covalent bonds and relativistic contractions. As documented in the literature, restricting excitations to the $6s^26p$ subspace, results in the loss of ``about half of the attraction is lost'' in Tl dimer interactions, whereas the explicit inclusion of the $5d$ shell in a 13-electron model recovers energetics that are nearly indistinguishable from the full 21-electron reference\cite{pyykko1999}. 
Despite this major reduction in variational freedom and the inherent loss of attraction in the 3-electron space, the large-core ccECP extracts the maximum possible accuracy from this restricted formulation. It maintains exceptional atomic fidelity (LMAD $<43$~meV) and transfers seamlessly to the TlO environment (Fig. \ref{fig:TlO_3}), where it substantially outperforms legacy BFD and MWBSTU potentials; which exhibit discrepancies exceeding $0.5$~eV at $R_{e} \approx 2.1$~\AA). While the ccECP exhibits a slight underbinding of $\sim$0.1~eV in TlH system (Fig. \ref{fig:TlH_3}), a direct, consequence of freezing the $5d$ shell \cite{pyykko1999}; it offers an almost flat discrepancy profile compared to the legacy ECPs, leading to the preservation of overall molecular properties compared to the AE reference and serve as a highly reliable foundation for large-scale cluster or condensed-phase calculations where explicit 13- or 21-electron treatments remain computationally prohibitive.

\subsubsection{Po}
The polonium (Po) ccECP ($Z_{\text{eff}}=6$) treats the $6s^{2}6p^{4}$ valence configuration explicitly. Modeling this deep core-valence boundary necessitates rigorously optimized parameters for the non-local channels to yield an exceptionally accurate atomic spectrum, with MAD and LMAD values of $\sim$0.05~eV (Figs. \ref{fig:MAD_in_elements}--\ref{fig:LMAD_in_elements}).

\begin{figure*}[!htbp]
\centering
\begin{subfigure}{0.5\textwidth}
\includegraphics[width=\textwidth]{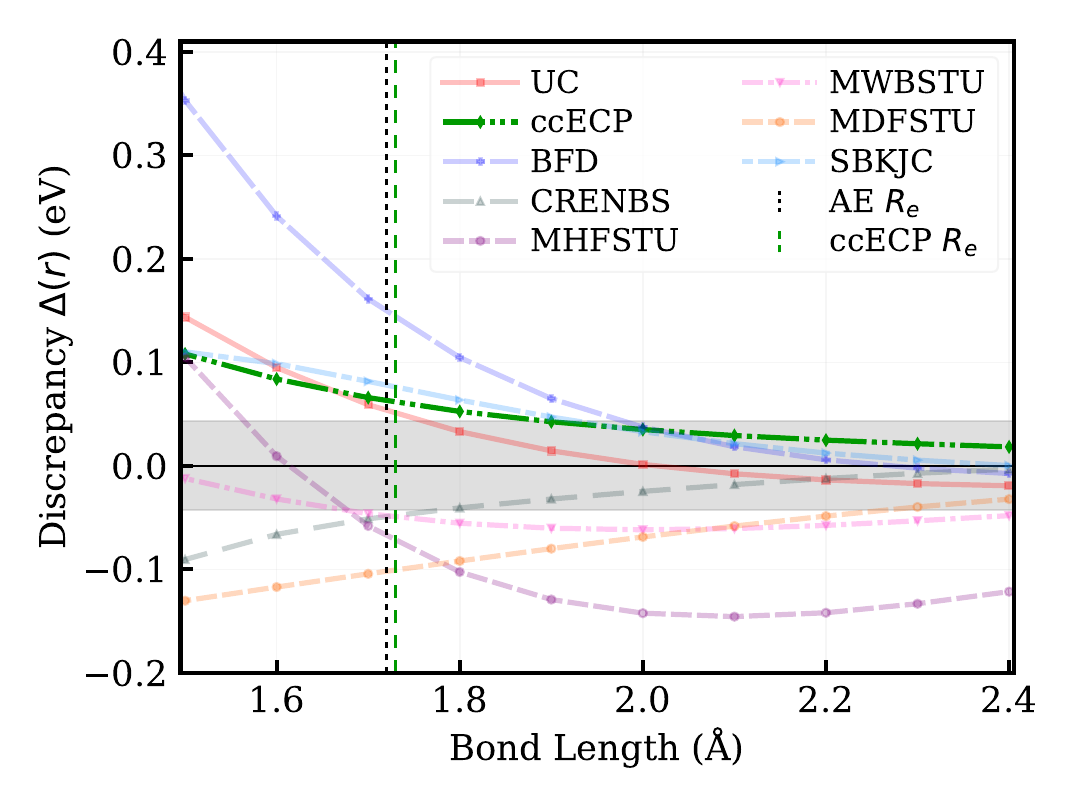}
\caption{CCSD(T) binding energy discrepancies for PoH molecule}
\label{fig:PoH}
\end{subfigure}%
\begin{subfigure}{0.5\textwidth}
\includegraphics[width=\textwidth]{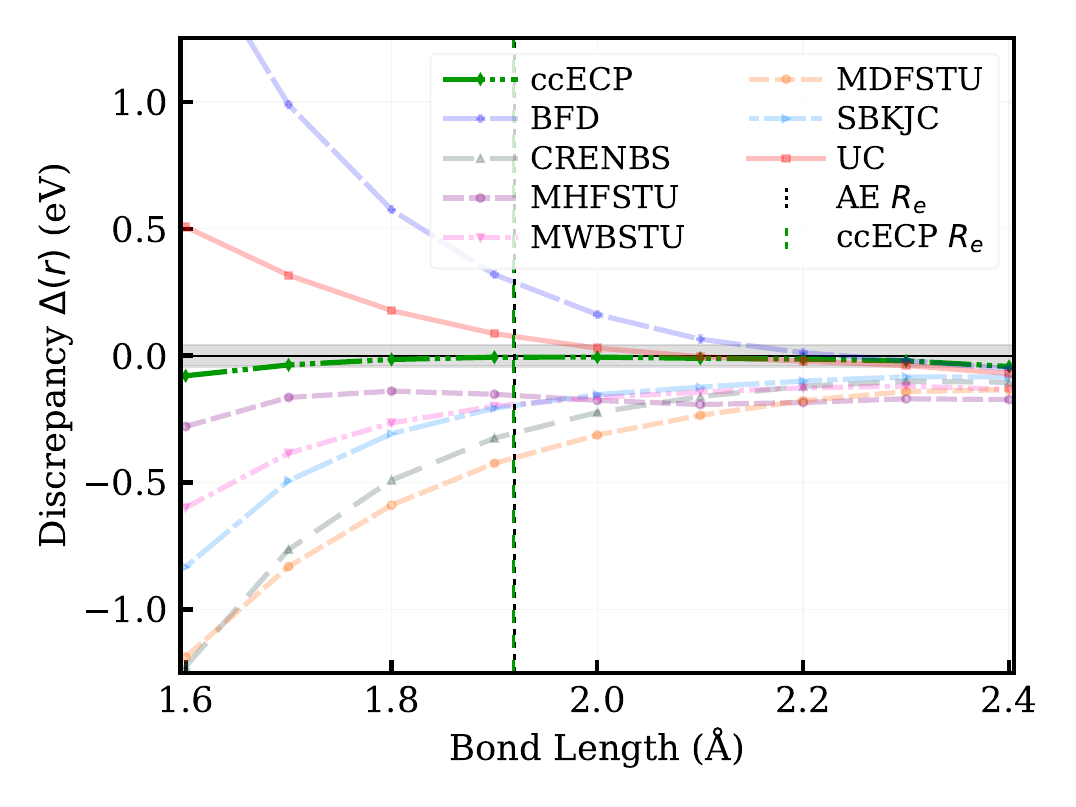}
\caption{CCSD(T) binding energy discrepancies for PoO molecule}
\label{fig:PoO}
\end{subfigure}
\caption{Binding energy discrepancies relative to the relativistic AE reference for (a) PoH and (b) PoO molecules. The graphical conventions are identical to those detailed in Fig. \ref{fig:Hf_mols}.
The ccECP is compared against available legacy ECPs that employ an equivalent 78-electron [[Xe]4$f^{14}$5$d^{10}$] core.} 
\label{fig:Po_mols}
\end{figure*}

This atomic precision seamlessly transfers to the PoO molecule (Fig. \ref{fig:PoO}), where the ccECP maintains a negligible discrepancy ($<43$~meV) across the entire potential energy surface, from compressed geometries to the dissociation limit. This provides a stark contrast to the performance of legacy 78-core ECPs, which exhibit significant deviations of the order of 0.5~eV at equilibrium and up to 1.0~eV in compressed regions. Historically, the literature has noted that legacy large-core pseudopotentials for heavy $6p$ elements often struggle to recover valence correlation energy and lack transferability compared to their small-core counterparts \cite{alsaidi2008, laury2012}. Consequently, previous studies on polonium have frequently abandoned 78-core potentials in favor of 60-core or AE treatments, noting that legacy large-core parameterizations often replaced too many core electrons and artificially constrained the core-valence boundary \cite{wodynski2016, zhutova2023}. However, the 78-core ccECP naturally reproduces the underlying scalar-relativistic physics without these traditional tradeoffs. Our predicted equilibrium bond length ($1.9189(8)$~\AA) is effectively indistinguishable from the rigorous AE relativistic benchmark ($1.9192(9)$~\AA), proving highly reliable for an open-shell species with significant thermodynamic sensitivity where capturing profound electron correlation is paramount \cite{mertens2019, van2015}.

We extend this precision to the PoH system (Fig. \ref{fig:PoH}), where bonding mechanics is heavily governed by prominent scalar-relativistic effects and intricate electron correlations rather than electrostatic interactions \cite{mertens2019, van2015}. Here, the ccECP demonstrates a marked improvement over legacy potentials such as MDFSTU and BFD, performing on par with the most accurate approximations in our test set. Most importantly, the ccECP maintains a remarkably flat discrepancy curve with no sharp slopes across the bonding regime, closely mirroring the AE reference. This stability directly translates into the derived spectroscopic properties as ccECP predicts an equilibrium bond length of $1.7295(29)$~\AA~ which is in excellent agreement with the respective AE value of $1.7203(27)$~\AA. Ultimately, ccECP successfully prioritizes the transferability required for reliable predictive modeling across diverse ionized states and complex chemical environments.

\begin{table*}
\small
\centering
\caption{SOREP optimized parameters for the selected $6p$ elements ccECPs. The parameters follow the same definitions as in Table \ref{tab:selected_5d_params}.
}
\label{tab:selected_5p_and_6p_params}
\begin{tabular}{cccccrrccccccrrr}
\hline
\hline
\multicolumn{1}{c}{Atom} & \multicolumn{1}{c}{$Z_{\rm eff}$} & \multicolumn{1}{c}{Hamiltonian} & \multicolumn{1}{c}{$\ell$} & \multicolumn{1}{c}{$n_{\ell k}$} & \multicolumn{1}{c}{$\alpha_{\ell k}$} & \multicolumn{1}{c}{$\beta_{\ell k}$} & & \multicolumn{1}{c}{Atom} & \multicolumn{1}{c}{$Z_{\rm eff}$} & \multicolumn{1}{c}{Hamiltonian} & \multicolumn{1}{c}{$\ell$} & \multicolumn{1}{c}{$n_{\ell k}$} & \multicolumn{1}{c}{$\alpha_{\ell k}$} & \multicolumn{1}{c}{$\beta_{\ell k}$} \\
\hline
   &    &      &   &   &             &               &&    &    &      &   &   &             &              \\
Tl & 21  & AREP & 0 & 2 &    9.687992 &  150.375457  && Tl & 13 & AREP & 0 & 2 &    3.940278 &   62.653492  \\    
   &     &      & 0 & 2 &    7.871228 &   62.089797  &&    &    &      & 0 & 2 &    1.621953 &   38.099901  \\
   &     &      & 1 & 2 &    9.598599 &   41.489210  &&    &    &      & 1 & 2 &    2.717680 &   55.807042  \\
   &     &      & 1 & 2 &    8.006580 &  159.651066  &&    &    &      & 1 & 2 &    1.061805 &   17.020907  \\
   &     &      & 2 & 2 &    7.320218 &   89.794632  &&    &    &      & 2 & 2 &    7.019022 &   69.601558  \\
   &     &      & 2 & 2 &    4.803356 &   25.920109  &&    &    &      & 2 & 2 &    2.084342 &   -4.950778  \\
   &     &      & 3 & 2 &    5.587719 &   36.929497  &&    &    &      & 3 & 2 &    6.195706 &   31.726110  \\
   &     &      & 3 & 2 &    2.848420 &    6.812020  &&    &    &      & 3 & 2 &    0.623552 &   -0.333353  \\
   &     &      & 4 & 1 &    9.936474 &   21.000000  &&    &    &      & 4 & 1 &    3.007525 &   13.000000  \\
   &     &      & 4 & 3 &    9.758692 &  208.665946  &&    &    &      & 4 & 3 &    2.986195 &   39.097825  \\
   &     &      & 4 & 2 &    9.208744 & -139.627859  &&    &    &      & 4 & 2 &    2.284716 &   -6.965301  \\
   &     &      & 4 & 2 &    9.006392 &   -0.599681  &&    &    &      & 4 & 2 &    1.792313 &  -14.937295  \\
   &     &      &   &   &             &              &&    &    &      &   &   &             &              \\
   &     & SO   & 1 & 2 &    7.454872 &   -9.325831  &&    &    & SO   & 1 & 2 &    0.912802 &   -5.221790  \\
   &     &      & 1 & 2 &    3.606950 &    9.432205  &&    &    &      & 1 & 2 &    0.860937 &    5.573587  \\
   &     &      & 1 & 2 &    9.045178 & -144.675271  &&    &    &      & 1 & 2 &    0.235977 &    0.318676  \\
   &     &      & 1 & 2 &    9.064520 &  144.473760  &&    &    &      & 1 & 2 &    0.114137 &   -0.104737  \\
   &     &      & 2 & 2 &    7.611101 &  -35.804179  &&    &    &      & 2 & 2 &    0.669406 &   -2.961710  \\
   &     &      & 2 & 2 &    6.509579 &   36.090203  &&    &    &      & 2 & 2 &    0.605691 &    2.822699  \\
   &     &      & 2 & 2 &    4.832549 &  -10.255748  &&    &    &      & 3 & 2 &    0.803350 &    1.432279  \\
   &     &      & 2 & 2 &    5.980170 &   10.496826  &&    &    &      & 3 & 2 &    0.791207 &   -1.330403  \\
   &     &      & 3 & 2 &    5.705058 &  -10.491867  &&    &    &      &   &   &             &              \\
   &     &      & 3 & 2 &    5.627611 &   10.493551  &&    &    &      &   &   &             &              \\
   &     &      & 3 & 2 &    2.953605 &   -1.884959  &&    &    &      &   &   &             &              \\
   &     &      & 3 & 2 &    2.900110 &    1.889992  &&    &    &      &   &   &             &              \\
   &     &      &   &   &             &              &&    &    &      &   &   &             &              \\
   &     &      &   &   &             &              &&    &    &      &   &   &             &              \\
Tl & 3  & AREP & 0 & 2 &    0.279913 &   -1.063522   && Po & 6  & AREP & 0 & 2 &    1.464230 &   33.828161  \\    
   &    &      & 0 & 2 &    2.104802 &   51.706270   &&    &    &      & 0 & 2 &    0.531139 &   -4.143278  \\
   &    &      & 1 & 2 &    0.478188 &   -2.966600   &&    &    &      & 1 & 2 &    1.790206 &   34.990194  \\
   &    &      & 1 & 2 &    1.005905 &   19.730111   &&    &    &      & 1 & 2 &    1.400424 &   -4.138807  \\
   &    &      & 2 & 2 &    0.432923 &    2.792410   &&    &    &      & 2 & 2 &    0.900710 &    5.485879  \\
   &    &      & 2 & 2 &    0.777879 &   -3.969764   &&    &    &      & 2 & 2 &    0.791500 &    7.936031  \\
   &    &      & 3 & 2 &    0.938200 &   -7.211192   &&    &    &      & 3 & 2 &    1.067376 &   -2.637644  \\
   &    &      & 3 & 2 &    0.439934 &    2.794325   &&    &    &      & 3 & 2 &    1.025831 &   -3.274796  \\
   &    &      & 4 & 1 &    0.971977 &    3.000000   &&    &    &      & 4 & 1 &    2.060069 &    6.000000  \\
   &    &      & 4 & 3 &    0.864457 &    2.915932   &&    &    &      & 4 & 3 &    1.888536 &   12.360434  \\
   &    &      & 4 & 2 &    0.973073 &    0.800299   &&    &    &      & 4 & 2 &    2.079841 &  -12.708696  \\
   &    &      & 4 & 2 &    0.879944 &   -6.033531   &&    &    &      & 4 & 2 &    1.103910 &   -0.094388  \\
   &    &      &   &   &             &               &&    &    &      &   &   &             &              \\
   &    &      &   &   &             &               &&    &    & SO   & 1 & 2 &    0.865225 &  -24.515466  \\
   &    &      &   &   &             &               &&    &    &      & 1 & 2 &    0.729930 &   22.294942  \\
   &    &      &   &   &             &               &&    &    &      & 1 & 2 &    0.744380 &    2.779634  \\
   &    &      &   &   &             &               &&    &    &      & 1 & 2 &    0.434625 &   -2.560364  \\
   &    &      &   &   &             &               &&    &    &      & 2 & 2 &    0.916242 &   -5.484215  \\
   &    &      &   &   &             &               &&    &    &      & 2 & 2 &    0.864955 &    5.289295  \\
   &    &      &   &   &             &               &&    &    &      & 3 & 2 &    1.105181 &    3.759223  \\
   &    &      &   &   &             &               &&    &    &      & 3 & 2 &    0.274518 &   -0.232713  \\
   &    &      &   &   &             &               &&    &    &      &   &   &             &              \\
\hline
\hline
\end{tabular}
\end{table*}

\subsubsection{At}
The electronic structure of astatine (At) represents the extreme limit of the halogen group, where massive scalar-relativistic effects and SO coupling fundamentally redefine chemical bonding\cite{shee2018,casetti2022} and predictions for exhibiting a metallic character in the condensed phase\cite{andreas2013}. For the 78-core ($Z_{\text{eff}}=7$), ccECP exhibits remarkable consistency, narrowing the discrepancy margin across all test sets. Atomic benchmarks (Figs. \ref{fig:MAD_in_elements}--\ref{fig:LMAD_in_elements}) show a MAD of $\sim$0.04~eV, effectively reaching chemical accuracy and outperforming SBKJC and MDFSTU, which deviate by roughly $\sim$0.07~eV and $\sim$0.13~eV, respectively.

\begin{figure*}[!htbp]
\centering
\begin{subfigure}{0.5\textwidth}
\includegraphics[width=\textwidth]{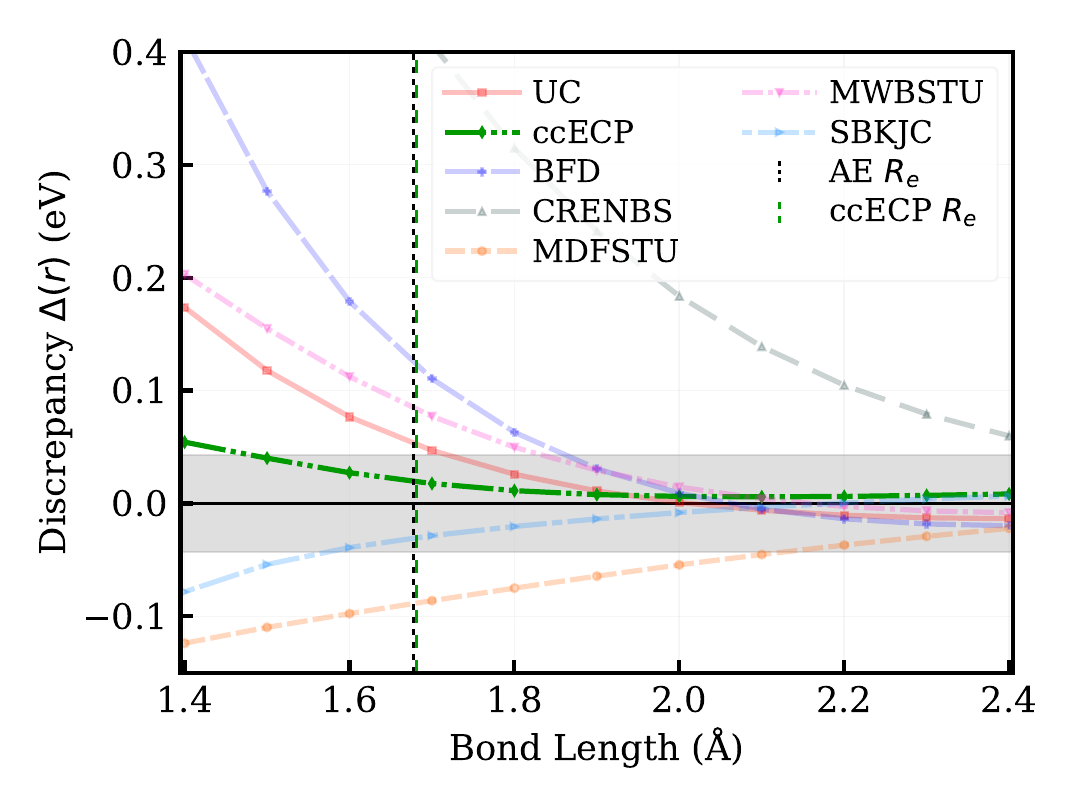}
\caption{CCSD(T) binding energy discrepancies for AtH molecule}
\label{fig:AtH}
\end{subfigure}%
\begin{subfigure}{0.5\textwidth}
\includegraphics[width=\textwidth]{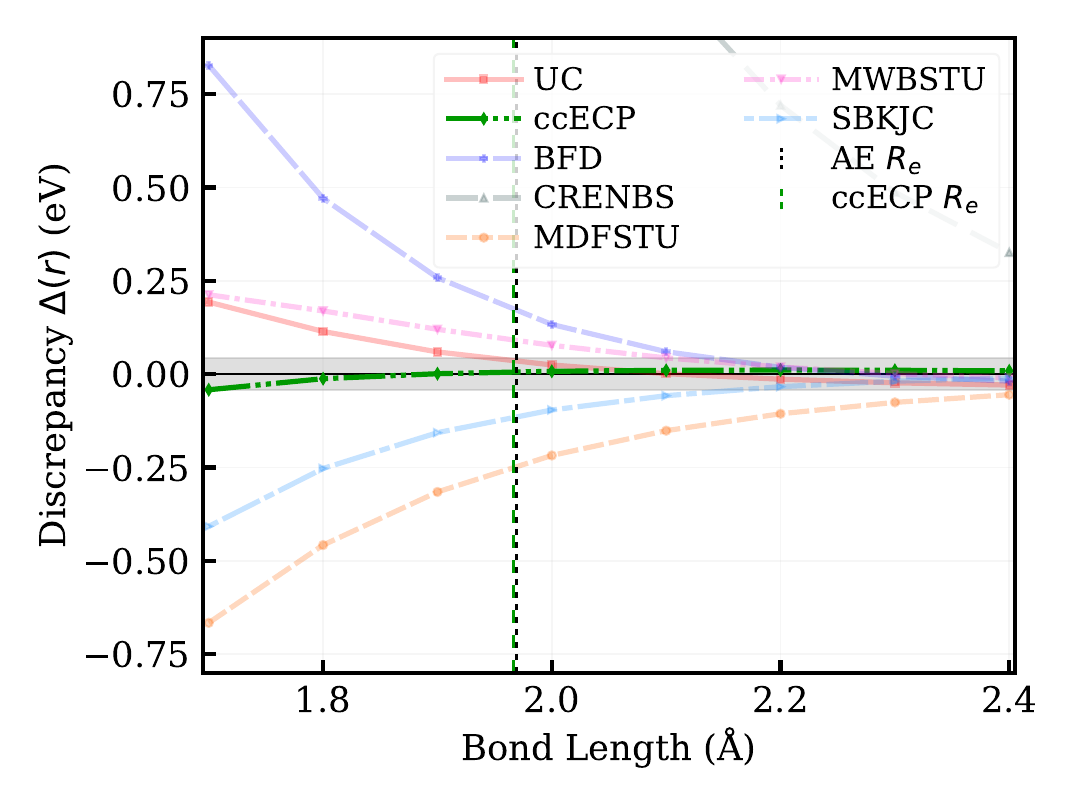}
\caption{CCSD(T) binding energy discrepancies for AtO molecule}
\label{fig:AtO}
\end{subfigure}
\caption{Binding energy discrepancies relative to the relativistic AE reference for (a) HAt and (b) AtO molecules. The graphical conventions are identical to those detailed in Fig. \ref{fig:Hf_mols}.
The ccECP is compared against available legacy ECPs that employ an equivalent 78-electron [[Xe]4$f^{14}$5$d^{10}$] core.}
\label{fig:At_mols}
\end{figure*}

This spectral accuracy translates into exceptional structural rigor in molecular environments. In the hydrogen astatide or astanine hydride (HAt/AtH) molecule (Fig. \ref{fig:AtH}), the ccECP successfully captures the pronounced relativistic contraction of the bond. Although less rigorous legacy approaches often underestimate this effect, leading to overestimated bond lengths, the ccECP yields an equilibrium bond length ($R_{e} \approx 1.681(3)$~\AA) in excellent agreement with our rigorous relativistic AE benchmark ($R_{e} \approx 1.678(2)$~\AA) which is consistent with the scalar-relativistic limit identified in previous studies\cite{casetti2022}. 

The broad transferability of the 78-core ccECP is most starkly demonstrated in the complex, open-shell AtO (Fig. \ref{fig:AtO}). Here, the ccECP maintains a negligible discrepancy ($<43$~meV) across the entire potential energy surface, from compressed geometries to the dissociation limit. In contrast, legacy 78-core ECPs exhibit significant deviations, on the order of 0.3~eV at equilibrium and up to 0.6~eV in compressed regions. Historically, capturing the intricate physics of astatine systems has proven exceptionally difficult; extreme scalar-relativistic effects and strong spin-orbit coupling cause traditional $\sigma$ and $\pi$ bonding paradigms to break down \cite{shee2018}. Consequently, the literature frequently notes that standard large-core approximations struggle to recover dynamic core-valence correlation and spatial polarization without explicitly including the $n=5$ shell \cite{dolg2000} or resorting to 60-core treatments \cite{casetti2022,demidov2024}. 

By maintaining a flat discrepancy curve, the 78-core ccECP provides a robust, scalar-relativistic foundation that seamlessly encapsulates these complex interactions. Its fidelity in modeling the open-shell character of AtO establishes a validated baseline for large-scale simulations where direct 4-component AE many-body calculations are computationally restrictive. Such high-accuracy transferability is essential to gain insight into superheavy element homology, where precise treatment of the relativistic valence shell is a strict prerequisite for qualitative and quantitative reliability \cite{demidov2024}.

\subsubsection{Rn}
Radon represents the heaviest experimentally known rare gas and is characterized by massive dipole polarizability that dictates its van der Waals properties\cite{runeberg1998}. The 78-core benchmarks reveal that the ccECP provides the most robust description of atomic and molecular properties among the tested ECPs. In the atomic spectrum (Figs. \ref{fig:MAD_in_elements}--\ref{fig:LMAD_in_elements}), the ccECP achieves a MAD of $\sim$0.05~eV and an LMAD of $\sim$0.04~eV, successfully reaching chemical accuracy and representing nearly a six-fold improvement over the BFD ECP. 

\begin{figure*}[!htbp]
\centering
\begin{subfigure}{0.5\textwidth}
\includegraphics[width=\textwidth]{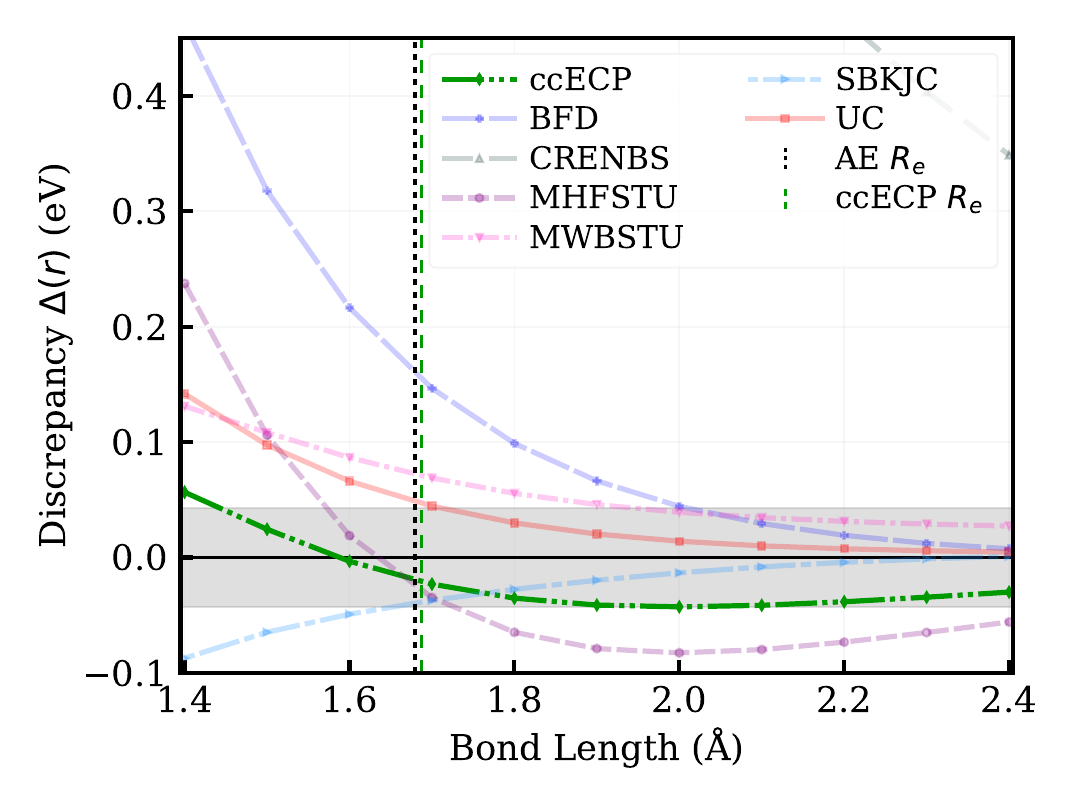}
\caption{CCSD(T) binding energy discrepancies for RnH$^{+}$ molecular ion}
\label{fig:RnH}
\end{subfigure}%
\caption{Binding energy discrepancies relative to the relativistic AE reference for RnH$^{+}$ cation. The graphical conventions are identical to those detailed in Fig. \ref{fig:Hf_mols}. The ccECP is compared against available legacy ECPs that employ an equivalent 78-electron [[Xe]4$f^{14}$5$d^{10}$] core.}
\label{fig:Rn_mols}
\end{figure*}

This atomic spectral fidelity is directly translated into molecular benchmarks for RnH$^{+}$ (Fig. \ref{fig:Rn_mols}). Here, the ccECP maintains a near-zero discrepancy ($<43$~meV) throughout the entire potential energy surface, almost indistinguishable from the AE benchmark. In contrast, legacy approximations exhibit significant deviations; the BFD ECP, for instance, deviates by $\sim$0.15~eV at the equilibrium bond length and up to $\sim$0.5~eV in compressed regions. Other legacy ECPs show similar discrepancies, demonstrating under or overbinding of the order of 0.1~eV in compressed geometries and failing to reach chemical accuracy at equilibrium. 

These historical deviations are deeply rooted in the physical complexity of radon. The literature extensively documents that radon exhibits massive dipole polarizability and deep core-valence correlation that dictate its chemical interactions \cite{nash2005}. Consequently, legacy large-core (78-electron) parameterizations was shown to yield flat potential energy surfaces \cite{lee1999} or required massive, uncontracted basis sets to recover missing dispersion and correlation \cite{runeberg1998}. To bypass these limitations, studies have often abandoned the 78-core approximations altogether in favor of 68- or 60-core potentials that explicitly correlate the $n=5$ shell \cite{fitzsimmons2010, nash2005}. By meticulously embedding these many-body correlation and scalar-relativistic effects directly into the non-local projectors, the 78-core ccECP resolves these historical shortcomings. It accurately captures profound relativistic effects without the computational bottleneck of an extended valence space.

Building upon this robust scalar baseline, the ccECP includes refined SO terms vital for theoretical exercises at the edge of the periodic table, where radon serves as the essential reference point for understanding the transition to superheavy chemistry in elements like oganesson \cite{nash2005,mitas2025}. Finally, ccECP addresses a significant practical gap in the available toolset for heavy 6p elements: because the MDFSTU library does not provide a [[Xe]$4f^{14}5d^{10}$] core SOREP, the ccECP offers a unique, highly reliable, and efficient large-core option for correlated many-body simulations of radon.

\begin{table*}
\small
\centering
\caption{SOREP optimized parameters for the selected $6p$ elements ccECPs. The parameters follow the same definitions as in Table \ref{tab:selected_5d_params}.
}
\label{tab:selected_5p_and_6p_params}
\begin{tabular}{cccccrrccccccrrr}
\hline
\hline
\multicolumn{1}{c}{Atom} & \multicolumn{1}{c}{$Z_{\rm eff}$} & \multicolumn{1}{c}{Hamiltonian} & \multicolumn{1}{c}{$\ell$} & \multicolumn{1}{c}{$n_{\ell k}$} & \multicolumn{1}{c}{$\alpha_{\ell k}$} & \multicolumn{1}{c}{$\beta_{\ell k}$} & & \multicolumn{1}{c}{Atom} & \multicolumn{1}{c}{$Z_{\rm eff}$} & \multicolumn{1}{c}{Hamiltonian} & \multicolumn{1}{c}{$\ell$} & \multicolumn{1}{c}{$n_{\ell k}$} & \multicolumn{1}{c}{$\alpha_{\ell k}$} & \multicolumn{1}{c}{$\beta_{\ell k}$} \\
\hline
   &    &      &   &   &             &              &&    &    &      &   &   &             &               \\
At & 7  & AREP & 0 & 2 &    7.287467 &  -15.175606  && Rn & 8  & AREP & 0 & 2 &    2.037762 &   35.773492   \\    
   &    &      & 0 & 2 &    2.527318 &   49.566653  &&    &    &      & 0 & 2 &    0.258894 &   -0.582176   \\
   &    &      & 1 & 2 &    1.403596 &   24.490846  &&    &    &      & 1 & 2 &    0.929092 &    8.364851   \\
   &    &      & 1 & 2 &    0.803040 &   -2.312971  &&    &    &      & 1 & 2 &    0.241082 &   -0.251247   \\
   &    &      & 2 & 2 &    1.454383 &   20.667048  &&    &    &      & 2 & 2 &    0.665009 &    2.954275   \\
   &    &      & 2 & 2 &    0.700867 &    5.266195  &&    &    &      & 2 & 2 &    0.696687 &    4.241122   \\
   &    &      & 3 & 2 &    1.155924 &   -2.685385  &&    &    &      & 3 & 2 &    0.785061 &   -2.150008   \\
   &    &      & 3 & 2 &    1.108198 &   -3.323071  &&    &    &      & 3 & 2 &    0.751270 &   -2.671971   \\
   &    &      & 4 & 1 &    4.000019 &    7.000000  &&    &    &      & 4 & 1 &    1.998502 &    8.000000   \\
   &    &      & 4 & 3 &    5.100037 &   28.000133  &&    &    &      & 4 & 3 &    2.107320 &   15.988014   \\
   &    &      & 4 & 2 &    3.999946 &   -0.500010  &&    &    &      & 4 & 2 &    1.983638 &  -15.107657   \\
   &    &      & 4 & 2 &    2.000303 &   -1.000050  &&    &    &      & 4 & 2 &    0.556948 &   -0.265576   \\
   &    &      &   &   &             &              &&    &    &      &   &   &             &               \\
   &    & SO   & 1 & 2 &    1.459087 &  -16.544521  &&    &    & SO   & 1 & 2 &    1.678230 &  -16.542491   \\
   &    &      & 1 & 2 &    2.064105 &   17.314288  &&    &    &      & 1 & 2 &    2.127395 &   17.287722   \\
   &    &      & 1 & 2 &    0.697770 &    3.367232  &&    &    &      & 1 & 2 &    0.844906 &    3.361482   \\
   &    &      & 1 & 2 &    1.633317 &   -0.845292  &&    &    &      & 1 & 2 &    1.626094 &   -0.857973   \\
   &    &      & 2 & 2 &    1.511891 &   -8.176153  &&    &    &      & 2 & 2 &    1.514288 &   -8.175977   \\
   &    &      & 2 & 2 &    1.449120 &    8.326947  &&    &    &      & 2 & 2 &    1.446023 &    8.327126   \\
   &    &      & 2 & 2 &    0.822625 &   -2.197131  &&    &    &      & 2 & 2 &    0.827483 &   -2.196374   \\
   &    &      & 2 & 2 &    0.760056 &    2.043230  &&    &    &      & 2 & 2 &    0.752937 &    2.044301   \\
   &    &      & 3 & 2 &    1.146909 &    1.791403  &&    &    &      & 3 & 2 &    1.243391 &    2.263688   \\
   &    &      & 3 & 2 &    1.095973 &   -1.662111  &&    &    &      & 3 & 2 &    0.881632 &   -1.215018   \\
   &    &      &   &   &             &              &&    &    &      &   &   &             &               \\
\hline
\hline
\end{tabular}
\end{table*}

\section{Conclusions}
\label{Conclusions}
This work completes the systematic development of the correlation-consistent effective core potentials library for the sixth row, filling for missing elements in $5d$ (Hf--Hg) and $6p$ (Tl--Rn) series. Predicting electronic structure properties of these heavy elements is notoriously challenging, as the competing scales of dynamic electron correlation and immense relativistic effects often cause standard legacy approximations to exhibit significant deviations. To address these competing demands, our construction strategy employs a physically rigorous partitioning of the core-valence boundary. For the $5d$ transition metals (Hf, Os, Hg) and the onset of the $6p$ block (Tl), we adopted a small-core ([[Kr]$4d^{10}4f^{14}$]) definition to explicitly capture the semi-core polarization and outer-core valence correlation frequently noted as essential in the literature. 

To further bridge the transition between these blocks, we introduced a highly accurate medium-core ([[Xe]$4f^{14}$]) variant for Hg and Tl. For all heavy $6p$ elements, a consistent large-core approach ([[Xe]$4f^{14}5d^{10}$]) was utilized. While legacy large-core parameterizations were less successful in reproducing effects of strong polarizability of these species, the rigorously optimized ccECP non-local projectors successfully absorb these complex scalar-relativistic interactions, maximizing computational efficiency without sacrificing valence accuracy. Crucially, the success of this intermediate-core resolution highlights a compelling path forward for broader pseudopotential design. Given that this intermediate core size tier was entirely bypassed in the lighter $3d/4p$ and $4d/5p$ rows, future work should retroactively explore comparable medium-core partitions for these lighter congeners, where significant room remains to unlock a finer, more optimized balance between computational throughput and many-body accuracy.

Distinct from many legacy approximations, the ccECP formalism explicitly regularizes the Coulomb singularity, enforcing a finite potential at the origin. This rigorous boundary condition ensures numerical stability and serves as a variance reduction technique in QMC applications avoiding thus necessity of effective nucleus cusps. Comprehensive benchmarking demonstrates that these ccECPs yield outstanding spectral accuracy, achieving a global average atomic mean absolute deviation of just $0.15$~eV which effectively halves the global discrepancies found in best legacy formulations ($\text{MAD}\approx0.31$~eV). Furthermore, the ccECPs constrain low-lying states MAD to a mere $0.045$~eV, resolving long-standing discrepancies in high-energy excitations and ionization potentials where legacy potentials showed mixed results. 

This atomic precision translates directly into robust molecular transferability across all three core partitions. Across the entire sixth row, the ccECPs systematically restrict $D_e$ discrepancies to under $0.03$~eV and constrain $R_e$ deviations to within $\sim$0.005~\AA. This represents an order-of-magnitude leap in reliability over legacy alternatives where $D_e$ deviations frequently exceed $0.16$~eV and discrepancies at the molecular dissociation limit ($\Delta D_\text{diss}$) deviate by up to $1.0$ to $3.0$~eV. By keeping $\Delta D_\text{diss}$ tightly bound below $0.10$~eV while delivering a uniformly excellent description of scalar-relativistic effects and spin-orbit coupling, this library removes a major methodological bottleneck, paving the way for predictive many-body simulations of complex heavy-element materials.

\section*{Supplementary Material}
\label{sec:suppl} 
The Supplementary Material provides comprehensive validation data, including scalar-relativistic atomic spectral discrepancies, molecular binding parameters derived from Morse fits, and spin-orbit multiplet splittings. It further details the Gaussian basis set generation protocols and assesses plane-wave energy cutoffs.

\section*{acknowledgments}
The authors thank Paul R. C. Kent for reading the manuscript and providing helpful suggestions.

This work has been supported by the U.S. Department of Energy, Office of Science, Basic Energy Sciences, Materials Sciences and Engineering Division, as part of the Computational Materials Sciences Program and Center for Predictive Simulation of Functional Materials.

This research used resources of the National Energy Research Scientific Computing Center (NERSC), a U.S. Department of Energy Office of Science User Facility operated under Contract No. DE-AC02-05CH11231.

An award of computer time was provided by the Innovative and Novel Computational Impact on Theory and Experiment (INCITE) program.

This research used resources of the Oak Ridge Leadership Computing Facility, which is a DOE Office of Science User Facility supported under Contract No. DE-AC05-00OR22725.

This paper describes objective technical results and analysis. Any subjective views or opinions that might be expressed in the paper do not necessarily represent the views of the U.S. Department of Energy or the United States Government.


\section*{Conflict of Interest}
The authors have no conflicts to disclose.

\section*{Author Contributions}
\textbf{Omar Madany}: Conceptualization (lead); Data Curation (lead); Investigation (lead); Methodology (lead); Validation (lead); Visualization (lead); Writing – original draft (lead); Writing – review \& editing (equal).
\textbf{Lubos Mitas}: Conceptualization (supporting); Investigation (supporting); Funding acquisition (lead); Project administration (lead); Resources (lead); Supervision (lead); Writing – review \& editing (equal).


\section*{Data Availability} 
The data supporting this study are available within the article and its Supplementary Material. The complete ccECP library, comprising semi-local parameters, non-local Kleinman-Bylander projectors, and basis sets in standard formats (e.g., \textsc{Molpro}, \textsc{DIRAC}, \textsc{NWChem}, \textsc{GAMESS}), are accessible at Ref.\cite{pseudopotentiallibrary}. Raw input/output archives are hosted by the Material Data Facility \cite{Blaiszik2016, Blaiszik2019} at Ref. \cite{mdf-5d-6p_data}.
\newpage
\clearpage
\section*{REFERENCES}
\bibliographystyle{unsrt}
\bibliographystyle{apsrev4-2}
\bibliography{main.bib}
\end{document}